\documentclass[final,3p,times]{elsarticle}

\usepackage{ae}
\usepackage{graphicx}
\usepackage{subfig}
\usepackage{amsmath}
\usepackage{amssymb}
\usepackage{amsfonts}
\usepackage{siunitx}
\usepackage{here}
\usepackage{tikz} \usetikzlibrary{shapes,arrows}
\usepackage{xcolor}
\usepackage{boldline}
\usepackage{multirow,array}
\usepackage{arydshln}
\usepackage{accents}
\usepackage{titlesec}
\begin{document}

\newcommand{\mat}[1]{{\boldsymbol{\mathbf{#1}}}}
\renewcommand{\vec}[1]{{\boldsymbol{\mathbf{#1}}}}

\newlength{\txtwt}
\setlength{\txtwt}{0.88\textwidth}
\newcommand{\txtw}{\textwidth}

\newcolumntype{P}[1]{>{\centering\arraybackslash}p{#1}}
\newcommand{\specialcellc}[2][c]{%
 \begin{tabular}[#1]{@{}c@{}}#2\end{tabular}}
\newcommand{\specialcelll}[2][c]{%
 \begin{tabular}[#1]{@{}l@{}}#2\end{tabular}}

\definecolor{myred}{RGB}{192,0,0}
\definecolor{myblue}{RGB}{0,112,192}
\definecolor{mygreen}{RGB}{0,176,80}
\definecolor{mygrey}{RGB}{201,201,201}

\newcommand{\CFL}{C\! F\! L}
\newcommand{\IBM}{\Gamma}
\newcommand{\f}{\mathrm{f}}
\renewcommand{\s}{\mathrm{s}}
\newcommand{\e}{\mathrm{e}}
\renewcommand{\d}{\;\mathrm{d}}

\makeatletter
\newcommand{\smallsym}[2]{#1{\mathpalette\make@small@sym{#2}}}
\newcommand{\make@small@sym}[2]{
  \vcenter{\hbox{$\m@th\downgrade@style#1#2$}}
}
\newcommand{\downgrade@style}[1]{
  \ifx#1\displaystyle\scriptstyle\else
    \ifx#1\textstyle\scriptstyle\else
      \scriptscriptstyle
  \fi\fi
}
\newcommand{\smallcirc}{\smallsym{\mathbin}{\circ}}
\newcommand{\smallvartriangle}{\smallsym{\mathbin}{\vartriangle}}
\newcommand{\smalltriangledown}{\smallsym{\mathbin}{\triangledown}}
\newcommand{\smallast}{\smallsym{\mathbin}{\ast}}
\makeatother
\newcommand{\qm}{\!\smallast\!}

\newcommand{\tra}{_\text{\tiny{0}}}
\newcommand{\trae}{_{\text{\tiny{0}},e}}

\renewcommand{\bar}{\overline}

\renewcommand{\fint}{\accentset{\smallvartriangle }{\vec{f}}}
\newcommand{\fext}{\accentset{\smalltriangledown}{\vec{f}}}
\newcommand{\mint}{\accentset{\smallvartriangle }{\vec{m}}}
\newcommand{\mext}{\accentset{\smalltriangledown}{\vec{m}}}

\titleformat*{\subsection}{\bfseries}
\titleformat*{\subsubsection}{\bfseries}

\setlength\parindent{0pt}

\begin{frontmatter}

\begin{flushright}
	\today
\end{flushright}
\vspace*{0.3cm}

\journal{Journal of Computational Physics}

\title{An immersed boundary method for the fluid-structure interaction of slender flexible structures in viscous fluid}

\author[PSM]{Silvio Tschisgale\corref{mycorrespondingauthor}}
\cortext[mycorrespondingauthor]{Corresponding author}
\ead{silvio.tschisgale@tu-dresden.de}

\author[PSM]{Jochen Fr\"ohlich}

\address[PSM]{Institut f\"ur Str\"omungsmechanik, Technische Universit\"at Dresden, George-B\"ahr Str. 3c, 01062 Dresden, Germany}

\begin{abstract}
This paper presents a numerical method for the simulation of fluid-structure interaction specifically tailored to interactions between Newtonian fluids and a large number of slender viscoelastic Cosserat rods. Because of their high flexibility and low weight the rods considered here exhibit large deflections, even under moderate fluid loads. Their motion, in turn, modifies the flow so that fluid and structures are strongly coupled to each other which is numerically very challenging. The paper proposes a new coupling approach based on an immersed boundary method which improves upon existing methods for this problem. It is numerically stable and exempt from any global iteration between the fluid part and the structure part, thus yielding high stability and low computational cost of the coupling scheme. The contribution presents the underlying methodology and its algorithmic realization, including an assessment of accuracy and convergence by systematic studies. Various validation cases illustrate performance and versatility of the proposed method.\\
\end{abstract}


\end{frontmatter}
\newpage

\section{Introduction}
\label{cha:introduction}

\subsection{Simulation of fluid-structure interaction}
During the last decades, various numerical approaches have been developed for the simulation of fluid-structure interactions (FSI). These numerical methods differ in the manner in which the fluid and the structure are coupled in time and space. The \textit{temporal coupling} can either be \textit{monolithic}, defining a single discrete system comprising the fluid and the structure, or \textit{partitioned}. With the partitioned approach, discrete equations for the fluid and the structure are solved separately and then coupled by an appropriate coupling algorithm. This strategy is versatile as it allows to use existing and optimized solvers. Hence, it is employed in most cases. The price to be paid is that the coupling can become unstable which requires substantial care and often generates problems. In this context, one can distinguish between \textit{weak} and \textit{strong} coupling strategies. With the former, also designated as \textit{explicit} coupling schemes, the fluid and the structural part are solved once within each time step with an exchange of coupling quantities, such as the instantaneous fluid loads on the structure, at the end of the step, for example. This exchange is often performed in a sequential manner which allows a simple implementation. With this approach, however, it is not guaranteed that the kinematic and dynamic coupling condition at the interface are fulfilled accurately. In addition, weak coupling schemes become numerically unstable if structures are mobile and lightweight, so that the added mass effect of the fluid becomes important~\cite{Forster2007, Markert2010, Huang2019}. Such kind of FSI problems require a strong coupling strategy also termed implicit coupling. Then, the fluid and solid part are usually solved repeatedly, iterating within each time step until the coupling condition at the interface satisfies a certain convergence criterion.\\
In addition to the temporal coupling of the fluid and the structure part, both need to be \textit{spatially coupled} at their common interface after discretization in space. The most common approach is to use a boundary fitted mesh to represent the structure within the fluid domain~\cite{Wall1998,Bungartz2006,Hartmann2008}. This simplifies the imposition of boundary conditions at the fluid-structure interface. But requires to adjust the grid in each time step, which is costly an can require additional measures to maintain grid quality. Furthermore, the grid in the fluid domain and the structure domain generally do not match, so that interpolation schemes have to be employed, rising issues of conservation properties and accuracy. Specific coupling software is often employed to implement these steps~\cite{MPCCI}. As an alternative to moving mesh techniques, approaches using a spatially uniform Eulerian background grid for the fluid part and a Lagrangian representation of the structures become increasingly popular~\cite{Mittal2005,Sotiropoulos2014}. This is due to various advantages over moving mesh methods like algorithmic simplicity, higher efficiency of the background fluid solver, etc. With a structure-independent, temporally constant fluid grid, the structures can be represented by various techniques, such as level-set (LS) methods \cite{Cottet2016}, volume-of-fluid (VOF) methods~\cite{Patel2017}, phase field (PF) methods \cite{Aland2013,Mokbel2018} or immersed boundary methods \cite{Sotiropoulos2014,Tian2014,Tullio2016,Kim2018}. Especially for simulations of flow through or around complex mobile geometries, the immersed boundary method (IBM) has been applied with great success during the past decade. Closely related is the so-called fictitious domain method~\cite{Glowinski1997} which was developed within the FEM framework. As stated in~\cite{Loon2007}, in the strong form the fictitious domain method does not differ from the immersed boundary method, but in the weak form when using an integral formulation of the FSI problem. Since these methods turned out to be well suited for scenarios with a large number of immersed mobile structures, e.g. particulate flows with thousands of particles~\cite{Vowinckel2014,Kidanemariam2017}, the IBM approach is used in this work as well. This is motivated by the ultimate goal of the present research to simulate scenarios comprising a large number of interacting even colliding slender structures. For such cases geometrically adapted grids for the fluid would be very difficult to devise and costly to employ.

\subsection{Immersed boundary methods for FSI problems}
The IBM was originally introduced by Peskin~\cite{Peskin1977}. Later on, a variety of different IB approaches were developed in recent years differing in various technical aspects as reviewed in~\cite{Mittal2005,Sotiropoulos2014}. While the fluid field is treated by an Eulerian description on a temporally constant fluid grid, the immersed structures are described using a Lagrangian point of view. In the general case, the grids of the movable structures do not conform with the fixed grid of the fluid. At this point, the IBM offers a method to impose the coupling conditions on the fluid-structure interface. Concerning the spatial imposition of the coupling conditions, IBMs are usually grouped into so-called \textit{discrete forcing} schemes and \textit{continuous forcing} schemes \cite{Mittal2005}. With the discrete forcing approach, the boundary conditions at the interface are imposed through the use of grid cells in the solid part. For each of these cells an interpolation scheme is derived that invokes the desired boundary condition at the interface~\cite{Kim2001,Mittal2005}. In the continuous forcing approach, compact delta functions are used at the interface for the transfer of quantities between the fluid and the immersed structures. A distinctive feature of a continuous forcing is, that the fluid-structure interface is represented by evenly distributed surface markers after spatial discretization of the physical problem~\cite{Uhlmann2005,Kempe2012}. This avoids the identification of special grid points for the imposition of the coupling conditions, with the drawback that the interface is ``smeared'' over several cells of the fluid grid, typically three to four cells around each of the marker points. In this region the local coupling force is introduced in the momentum balance of the fluid to impose the no-slip condition at the fluid-structure interface. Due do its simplicity, stability and high efficiency, IBMs with continuous forcing are used preferably in large-scale simulations, e.g. disperse multiphase flows with rigid particles~\cite{Vowinckel2014,Kidanemariam2017} or bubbles~\cite{Santarelli2015}.\\ 
Besides the different approaches used for the spatial coupling, the various IBMs differ in the manner in which the coupling force is computed in time, when a partitioned coupling approach is applied. Familiar techniques are \textit{feedback forcing}, \textit{discrete mass} and \textit{momentum forcing} as well as the so-called \textit{direct forcing} \cite{Prosperetti2009}. The direct forcing approach is one of the most popular methods because of its increased stability.\\
In the literature, several IBMs can be found with movable rigid bodies, e.g.~\cite{Glowinski2001,Gilmanov2005,Kempe2012,Sotiropoulos2014,Kim2016}. The description of fluid-structure interactions in the narrow sense, with elastic solid structures, is less common but has become increasingly important over the last decade. Most of these IBMs, however, were implemented and tested only with two-dimensional cases~\cite{Baaijens2001,Zhu2002,Yu2005,Loon2007,Richter2013,Favier2014,Verkaik2015}. Recently, more and more efforts have been made to simulate truly three-dimensional scenarios. These can be divided into fluid-structure interactions with one-dimensional fiber-like structures~\cite{Griffith2012,Bhalla2013,Wiens2015-2}, two-dimensional elastic membranes~\cite{Le2009,Zhu2014,Wiens2015-1,Tullio2016} and volumetric elastic structures~\cite{Zhang2004,Tian2014,Kim2018}. In some of these implementations, non-classical structure models are used, such as neutrally buoyant fibers and membranes in Le~\textit{et al.}~\cite{Le2009}, Griffith and Lim~\cite{Griffith2012}, Bhalla~\textit{et al.}~\cite{Bhalla2013} and Wiens and Stockie~\cite{Wiens2015-1,Wiens2015-2}, or a mass-spring network model in the work of de Tullio and Pascazio~\cite{Tullio2016}. A classical continuum mechanical description of the structures was applied by Zhang~\textit{et al.}~\cite{Zhang2004}, Tian~\textit{et al.}~\cite{Tian2014}, Zhu~\textit{et al.}~\cite{Zhu2014}, Gilmanov~\textit{et al.}~\cite{Gilmanov2015, Gilmanov2018} and recently by Kim~\textit{et al.}~\cite{Kim2018}. The methods mentioned, including IBMs for two-dimensional problems, cover a variety of coupling algorithms. Besides a few monolithic schemes~\cite{Baaijens2001,Loon2007,Richter2013,Verkaik2015}, most of the implementations are realized by means of a partitioned coupling approach. The latter range from non-iterative coupling schemes~\cite{Zhu2002,Zhang2004,Yu2005,Griffith2012,Favier2014,Zhu2014,Wiens2015-1} and iterative strong coupling approaches~\cite{Bhalla2013,Tian2014,Tullio2016,Gilmanov2015}, to improved non-iterative schemes with extended numerical stability~\cite{Le2009,Kim2018}. Sotiropoulos and Yang~\cite{Sotiropoulos2014} provided a comprehensive overview of various IB approaches for the simulation of general FSI problems distinguishing between weak and strong coupling strategies. An even more recent review of IBMs for fluid-structure interactions was published by Kim and Choi~\cite{Kim2019}.

\subsection{Basic idea of the coupling approach}
The IBM developed in this work can be assigned to the group of IBMs with continuous direct forcing. A special component of this coupling scheme is a novel non-iterative semi-implicit direct forcing which combines the stability of monolithic methods with the advantages of partitioned weak approaches. Furthermore, a general coupling strategy is proposed to couple of the Navier-Stokes equations with an arbitrary immersed structure, demonstrated for Cosserat rods here.\\
In contrast to other non-iterative coupling strategies, the main idea is not based on a stabilization technique, e.g. a relaxation technique~\cite{Dettmer2013,Tian2014,Tullio2016,Kim2018,Kadapa2018}.
Even if relaxation techniques are easy to implement, may offer numerical stability as well as a second order accuracy~\cite{Dettmer2013}, they also have their disadvantages. As mentioned in~\cite{Kadapa2018}, problems which feature strong added mass effects require small values of the relaxation parameter to obtain stability which, in turn, causes higher truncation errors and small time step sizes. The present direct forcing approach does not require any additional parameter. Numerical stability is achieved by using a semi-implicit time scheme for the structure motion. As a matter of fact, the coupling terms used to impose the coupling conditions, require some kind of implicit integration in time to ensure numerical stability~\cite{Fernandez2006}. It is shown here, that this is not only feasible by means of a global iteration between the fluid and structure part, but can also be achieved by an implicit integration of the coupling terms in the structure equations. These coupling terms are provided in a temporally continuous form so that the coupled structure equations can be discretized in time by an arbitrary implicit integration scheme. As a result, the FSI coupling becomes independent of the discretization techniques employed for both subsolvers, lending itself to application in a broad set of conditions. The scheme developed is completely non-iterative and requires only a single bidirectional exchange of coupling quantities between the fluid solver and the structure solver. Its only drawback is a first-order accuracy of the coupling terms in time which will be discussed in detail.

\section{Physical model and governing equations}
\label{cha:phyical_model}

\subsection{Problem definition and assumptions}
\label{sec:problem_def}
\begin{figure}[!tb]
  \centering
  \includegraphics[trim=0 0 0 0, clip, scale=1]{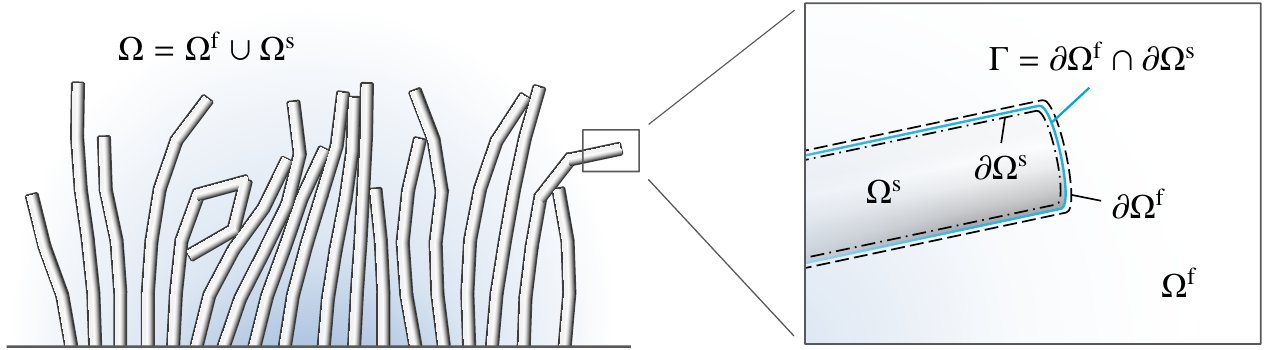}
	\caption{\label{fig:FSI_setup}Type of fluid-structure interaction considered in the present work, namely the interaction between viscous fluids and a larger number of slender flexible structures. The fluid domain and structure domain are designated as $\Omega^\f$ and $\Omega^\s$, respectively, with the corresponding boundaries $\partial\Omega^\f$ and $\partial\Omega^\s$. The fluid-structure interface is referred to as $\Gamma$.}
\end{figure}

The physical configuration addressed here consists of a viscous fluid interacting with a large number of flexible structures. These are assumed to be long and slender, as encountered with a fiber suspension or a canopy flow, as illustrated in Fig.~\ref{fig:FSI_setup} for example. Constant material properties are assumed for fluid and structures. All structures are assumed completely immersed in the fluid and are geometrically characterized by a long and slender shape with cross sections much smaller than their longitudinal extension. When the structures are subjected to fluid loads, local deformations and associated internal strains are assumed to be small but may agglomerate to large overall displacements in space.\\
The domain of the entire physical configuration $\Omega\in\mathbb{R}^3$ consists of the closed subset $\Omega^\f\subset\Omega$, defining the fluid domain, and a certain number of structures $N_\s$, which combine to form the closed subset $\Omega^\s\subset\Omega$, the structure domain, so that the union of the fluid domain and the set of all structures gives the entire domain $\Omega=\Omega^\f \cup \Omega^\s$ assumed to be time-independent here. The fluid domain $\Omega^\f$ and the structure domain $\Omega^\s$ may change their shape in time. The associated boundaries of both subdomains are $\partial^\f\Omega\subset\Omega^\f$ and $\partial\Omega^\s\subset\Omega^\s$, respectively, so that their intersection defines a time-dependent fluid-structure interface $\Gamma=\partial\Omega^\f\cap\partial\Omega^\s$. The boundary of the entire domain is given by $\partial\Omega=(\partial\Omega^\f\cup\partial\Omega^\s)\setminus\Gamma$.

\subsection{Individual models for fluid and structure}
\label{sec:fluidmech}

\subsubsection{Navier-Stokes equations}
The governing equations for the fluid motion are the unsteady three-dimensional Navier-Stokes equations (NSE) for a Newtonian fluid of constant density
\begin{subequations}
\label{eq:NSE}
\begin{align}
	\label{eq:IGL}
	\frac{\partial{\vec{u}}}{\partial{t}} + \nabla\cdot(\vec{u}\otimes\vec{u}) &= \frac{1}{\rho_\f}\:\nabla\cdot\mat{\sigma}+\vec{f} \\
	\label{eq:KGL}
	\nabla\cdot\vec{u}&=0
\end{align}
\end{subequations} 
in the fluid domain $\Omega^\f$, where $\vec{u}=(u,v,w)^\top$ designates the velocity vector in Cartesian components along the Cartesian coordinates $x,y,z$, while $t$ represents the time, $p$ the pressure field, and $\rho_\f$ the fluid density. The hydrodynamic stress tensor $\mat{\sigma}$ is defined by
\begin{equation}
	\label{eq:stress_tensor}
	\mat{\sigma} = -p \:\mathbb{I} + \mu_\f \:(\,\nabla \vec{u} + \nabla \vec{u}^\top ) \, ,
\end{equation}
with $\mu_\f\!=\!\rho_\f \, \nu_\f$ the dynamic viscosity and $\nu_\f$ the kinematic viscosity, $\mathbb{I}$ the identity matrix, and $\vec{f}\!=\!(f_{x},f_{y},f_{z})^\top$ a mass-specific force. The latter consists of two parts, \text{$\vec{f}=\vec{f}_V + \vec{f}_\IBM$}, where $\vec{f}_V$ is a mass-specific volume force, e.g. gravitational acceleration, and $\vec{f}_\IBM$ a coupling force used to impose the no-slip condition on the fluid-structure interface $\Gamma$, as described in section~\ref{cha:cfsi} below.

\subsubsection{Geometrically exact Cosserat rod model}
\label{sec:exact Cosserat rod model}
\begin{figure}[!tb]
	\centering
  \includegraphics[trim=0 0 0 0, clip, scale=1]{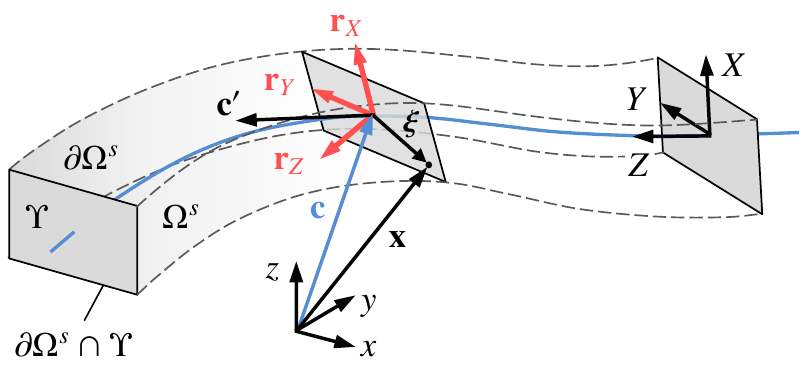}
	\caption{\label{fig:Cosserat_setup}Instantaneous configuration of a deformed Cosserat rod \,$\Omega^s$. Each of the cross sections $\Upsilon(Z)\subset\Omega^s$ along the arc length $Z$ remains rigid and plane, while its orientation is described by three orthonormal direction vectors. The latter can be combined to the rotation matrix $\mat{R}=[\vec{r}_X\,\vec{r}_Y\,\vec{r}_Z]$. The location of a material point $\vec{x}\in\Omega^s$ can be decomposed into a longitudinal part along $\vec{c}$ and a lateral part $\vec{\xi}=\mat{R}\cdot\vec{\xi}\tra$ within the associated cross-section $\Upsilon$, so that $\vec{x} = \vec{c}+\mat{R}\cdot\vec{\xi}\tra$.}
\end{figure}

The structures addressed here are characterized by a long and slender shape with cross sections much smaller than their longitudinal expansion. This kind of structure is usually referred to as a~\textit{beam},~\textit{cantilever} or~\textit{rod}, the latter term being used in the present work. In principle, such geometrical constraints on shape can be used to employ model reduction techniques which reduce the degrees of freedom required to describe the structure motion. These techniques are of crucial importance when simulating large numbers of individual resolved rod structures due to the enormous reduction of computational effort this entails. Especially for the slender rods considered here, the general three-dimensional equations of motion are well approximated by one-dimensional rod models without loss of physical correctness. One of the most complex rod models is the so-called~\textit{geometrically exact Cosserat rod} which covers both the rigid body motion and the common deformation modes of a rod~\cite{Simo1985, Antman1995, Auricchio2008, Lang2011}. This model is used in the present work, since it captures large structural displacements to be considered here, and offers a broad range of applications. Geometry and coordinate systems used for the definition of the Cosserat rod are assembled in Fig.~\ref{fig:Cosserat_setup}.  With this model each cross section is assumed to remain rigid during deformation (\textit{Euler--Bernoulli hypothesis}~\cite{Bauchau2009}), while internal strains are measured by the relative position and orientation between adjacent cross sections. On the basis of this kinematic constraint, the three-dimensional linear and angular momentum balance can be transferred into two spatially one-dimensional differential equations for the rod motion. One equation describes the temporal evolution of $\vec{c}(Z,t)\in\zeta$, i.e. the positions of the center line $\zeta\subset\Omega^\s$ of the rod. The second equation describes the rotation of the cross sections $\Upsilon(Z)\subset\Omega^\s$ along the Lagrangian arc length coordinate $Z$, captured by the rotation matrix $\mat{R}(Z,t)\in SO(3)$ (Fig.~\ref{fig:Cosserat_setup}). 
This equation is not formulated directly for the rotation matrix $\mat{R}(Z,t)$, but for the angular velocity $\vec{\omega}(Z,t)$ of the cross section. Both quantities are related via
\begin{equation}
	\label{eqn:relation_omega_R}
	[\vec{\omega}]_\times = \dot{\mat{R}} \cdot \mat{R}^\top ,
\end{equation}
where $[\vec{\omega}]_\times$ is the skew matrix of $\vec{\omega}$, so that $[\vec{\omega}]_\times\cdot\vec{v}= \vec{\omega}\times\vec{v}$ for any vector $\vec{v}\in\mathbb{R}^3$. 
The final form of the equations of motion, the so-called (geometrically exact) Cosserat rod equations~\cite{Simo1985, Antman1995, Lang2011}, reads
\begin{subequations}
	\label{eqn:Cosserat}
	\begin{align}
		\label{eqn:Cosserat_linear}
		\rho_\s A \, \ddot{\vec{c}} &= \fint' + \fext \\
		\label{eqn:Cosserat_angular}
		\rho_\s\mat{I}\!\cdot\!\dot{\vec{\omega}}\,+\,\vec{\omega}\times\rho_\s\mat{I}\!\cdot\!\vec{\omega} &= \mint' + \vec{c}'\times\fint + \mext \, ,
	\end{align}
\end{subequations}
where the temporal and spatial derivatives are abbreviated as $\dot{\vec{c}}=\mathrm{d}\vec{c}/\mathrm{d} t$, $\ddot{\vec{c}}=\mathrm{d}^2\vec{c}/\mathrm{d} t^2$ and $\vec{c}'=\partial\vec{c}/\partial Z$, respectively. In the present work, the rods have spatially constant geometrical properties, i.e. a constant cross sectional area $A$ and tensor of inertia $\mat{I}\tra$, as well as constant material properties, such as the density $\rho_\s$.\\
The motion of the rods, governed by Eqs.~\eqref{eqn:Cosserat}, depends on the internal forces $\fint$ and internal moments $\mint$, as well as on the external forces $\fext$ and external moments $\mext$. The external loads contain gravitational forces $\fext_\mathrm{g} = (\rho_\s\!-\!\rho_\f)\, A \,\vec{g}$, and external fluid loads acting on the fluid-structure interface $\Gamma$, denoted as $\fext_\IBM$ and $\mext_\IBM$. The internal forces $\fint_\IBM$ and moments $\mint_\IBM$ in Eqs.~\eqref{eqn:Cosserat} are formulated for a linear viscoelastic material of Kelvin-Voigt type~\cite{Meyers1999,Lang2011}, i.e.
\begin{subequations}
	\label{eqn:internal_loads}
	\begin{align}
		\label{eqn:fint}
		\fint &= \mat{C}_\varepsilon\cdot\left(\vec{\varepsilon}\,-\,\vec{\varepsilon}|_{t=0}\right) \,+\, \mat{C}_{\dot{\varepsilon}}\cdot\dot{\vec{\varepsilon}} \\
		\label{eqn:mint}
		\mint &= \mat{C}_\kappa\cdot\left(\vec{\kappa}-\vec{\kappa}|_{t=0}\right) \,+\, \mat{C}_{\dot{\kappa}}\cdot\dot{\vec{\kappa}} \, .
	\end{align}
\end{subequations}
Here, the first term on the right-hand side of Eq.~\eqref{eqn:fint} and~\eqref{eqn:mint} constitutes the linear elastic part of $\fint$ and $\mint$. Internal strains generated during deformation are measured by the strain vector $\vec{\varepsilon}$ and the curvature vector $\vec{\kappa}$. Both vectors are defined by~\cite{Lang2011}
\begin{equation}
	\label{eqn:strains}
	\vec{\varepsilon} = \vec{c}' \qquad \text{and} \qquad [\vec{\kappa}]_\times = \mat{R}' \, ,
\end{equation}
respectively. Their linear relation to the internal loads $\fint$ and $\mint$ are represented by the two constitutive matrices $\mat{C}_\varepsilon \!= \mat{R}\cdot\mat{C}_{\varepsilon_0}\cdot\mat{R}^\top$ with $\mat{C}_{\varepsilon_0} \!= \text{diag}\!\left(k_{\mathrm{s}_1}G_\s,\,k_{\mathrm{s}_2}G_\s,\,E_\s \right) A$, and $\mat{C}_\kappa \!= \mat{R}\cdot\mat{C}_{\kappa_0}\cdot\mat{R}^\top$ with $\mat{C}_{\kappa_0} \!= \text{diag}\!\left(E_\s,\,E_\s,\,k_{\mathrm{t}}G_\s\right)\mat{I}\tra$ where $E_\s$ is the Young modulus and $G_\s$ the shear modulus. Here, the geometric tensor of inertia \text{$\mat{I}\tra \!= \text{diag}({I_X,I_Y,J})$} contains the second moments of area $I_X$, $I_Y$ and $J=I_Z$ around the $\vec{r}_X$-, $\vec{r}_Y$- and $\vec{r}_Z$-axis of the cross section, respectively. The shear and torsion correction factors $k_{\mathrm{s}_1}$, $k_{\mathrm{s}_2}$ and $k_{\mathrm{t}}$ are used to model the influence of warping effects in case of shear and torsional loads.\\
The second term on the right-hand side of Eq.~\eqref{eqn:internal_loads} takes into account the dissipative part of the internal loads $\fint$ and $\mint$ due to internal friction. For the Kelvin-Voigt material used here, these depend linearly on the strain rate $\dot{\vec{\varepsilon}}$ and the curvature rate $\dot{\vec{\kappa}}$, respectively, while the corresponding constitutive matrices are given by $\mat{C}_{\dot{\varepsilon}} = \mat{R}\cdot\mat{C}_{\dot{\varepsilon}_0}\cdot\mat{R}^\top$ with $\mat{C}_{\dot{\varepsilon}_0} = \text{diag}(c_{\mathrm{s1}},c_{\mathrm{s2}},c_\mathrm{e})$, and $\mat{C}_{\dot{\kappa}} = \mat{R}\cdot\mat{C}_{\dot{\kappa}_0}\cdot\mat{R}^\top$ with $\mat{C}_{\dot{\kappa}_0} = \text{diag}(c_\mathrm{b1},c_\mathrm{b2},c_\mathrm{t})$. The subscript of each damping parameter~$c$ denotes the deformation mode, i.e. shear, extension, bending and torsion.

\subsection{Fluid-structure coupling}
\label{sec:fsi_coupling}
\subsubsection{Coupling conditions}
\label{subsec:coupling_condition}
The coupling between the Navier-Stokes equations~\eqref{eq:NSE} and the Cosserat rod equations~\eqref{eqn:Cosserat} is realized by the dynamic and the kinematic coupling condition. Both are applied at the common fluid-structure interface~$\Gamma$. The dynamic condition states the equality of the stress vectors, i.e.
\begin{equation}
  \label{eqn:dynamic_condition}
	\mat{\sigma}\cdot\vec{n}=\mat{\sigma}_\s\cdot\vec{n} \qquad \forall\vec{x}\in\Gamma \, , 
\end{equation}
with the hydrodynamic stress tensor $\mat{\sigma}$ according to~\eqref{eq:stress_tensor}, the structural Cauchy stress tensor $\mat{\sigma}_\s$, and the unit normal vector $\vec{n}$ pointing from the fluid domain $\Omega^\f$ into the solid domain $\Omega^\s$. The stress vector can be interpreted as a surface-specific coupling force $\vec{f}_S=\mat{\sigma}\cdot\vec{n}$ connecting both parts of the coupled system at the interface $\Gamma$. To impose the dynamic coupling condition~\eqref{eqn:dynamic_condition}, $\vec{f}_S$ has to be introduced in the momentum balances of the fluid and the structure. For the one-dimensional Cosserat rod considered here, the coupling force $\fext_\IBM(Z,t)$ and the coupling moment $\mext_\IBM(Z,t)$ read
\begin{equation}
	\label{eqn:fluid_loads_3D}
	\fext_\IBM=-\int_{\Gamma\cap\Upsilon}\vec{f}_S \d C \,, \qquad \mext_\IBM=-\int_{\Gamma\cap\Upsilon}\vec{x}\times\vec{f}_S \d C \, ,
\end{equation}
respectively. The kinematic coupling is realized by the no-slip boundary condition
\begin{equation}
  \label{eqn:no-slip_condition}
	\vec{u}=\vec{v} \quad \forall\vec{x}\in\Gamma \, ,
\end{equation}
which for the Cosserat rod with $\vec{v} = \dot{\vec{x}}$ and $\vec{x} = \vec{c}+\mat{R}\cdot\vec{\xi}\tra$ yields
\begin{equation}
  \label{eqn:no-slip_condition2}
	\vec{u}=\dot{\vec{c}} + \vec{\omega}\times \vec{\xi} \quad \forall\vec{x}\in\Gamma \, .
\end{equation}

\subsubsection{Zero-thickness assumption and jump conditions}
\label{sec:zero_thickness}
\begin{figure}[!tb]
  \centering
  \includegraphics[trim=0 0 0 0, clip, scale=1]{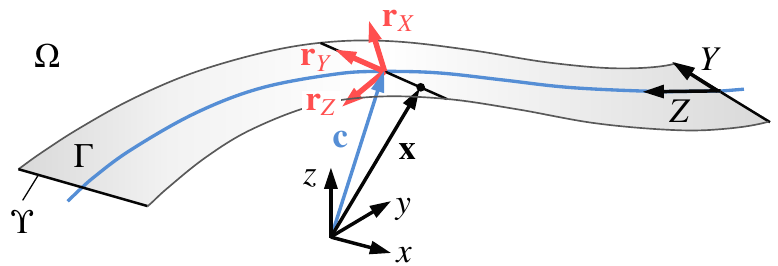}
  \caption{\label{fig:Cosserat_setup_zero}Representation of a strip-shaped rod in the fluid domain $\Omega^\f=\Omega$ as an infinitely thin Cosserat rod $\Omega^\s=\Gamma$. In contrast to the three-dimensional volumetric rod shown in Fig.\ref{fig:Cosserat_setup}, the cross sections $\Upsilon(Z)$ are one-dimensional and a subset of the fluid-structure interface $\Gamma$. 
 }
\end{figure}

\begin{figure}[!tb]
  \centering
	\includegraphics[trim=0 0 0 0, clip, scale=1]{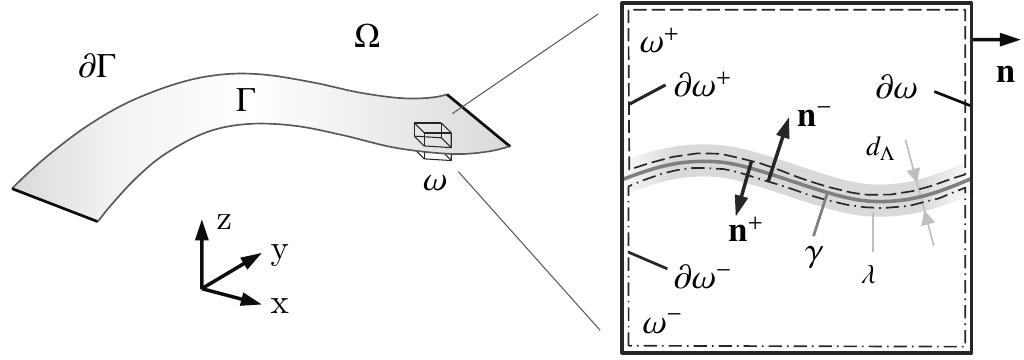}
  \caption{\label{fig:domain_definition}Illustration of an \textit{interface problem} in which the fluid domain $\Omega$ is cut by the interface $\Gamma$. On a smaller scale, the finite volume $\omega\subset\Omega$ is separated into two subdomains $\omega^+$ and $\omega^-$. The intersection of their boundaries $\partial\omega^+$and $\partial\omega^-$ defines the common interface $\gamma\subset\Gamma$. The vectors $\vec{n}$, $\vec{n}^+$ and $\vec{n}^-$ are unit normal vectors at the boundaries $\partial\omega$, $\partial\omega^+$and $\partial\omega^-$. The gray shaded area around $\gamma$ represents a compact, volumetric layer $\lambda$ of infinitesimally small thickness $d_\Lambda\to 0$, as described in section~\ref{sec:dist source}. }
\end{figure}

The long slender rods considered in this work have cross-sectional expansions much smaller than their longitudinal expansion. For the applications below, the rods are very well represented in the fluid as simple one-dimensional curves, e.g. to model fibers, or as two-dimensional geometries in case of strip-shaped rods (Fig.~\ref{fig:Cosserat_setup_zero}). Since at least one lateral expansion of the structure is neglected, the approach is often referred to as \textit{zero-thickness approximation} which is widely used in the literature to model rods or membranes~\cite{Bungartz2006,Dettmer2006,Huang2007,Lee2015,Zhu2014,Tian2014,Tullio2016}. This strategy is pursued here as well. The FSI model is tailored to rectangular cross sections of width $W$ and thickness $T$, with an aspect ratio of $T \ll W$. While the Cosserat-rod equations~\eqref{eqn:Cosserat} are solved for the three-dimensional volumetric rod with $T>0$, the rod is represented in the fluid as a two-dimensional object with $T = 0$, as illustrated in Fig.~\ref{fig:Cosserat_setup_zero}. 
When applying the zero-thickness assumption in the fluid domain, the structure domain $\Omega^\s$ completely coincides with the fluid-structure interface $\Gamma$, so that $\Omega^\s = \Gamma$. Moreover, the fluid domain now represents the entire domain of the coupled problem, i.e. $\Omega=\Omega^\f\cup\Omega^\s=\Omega^\f$, and contains the structure as an embedded fluid-structure interface $\Gamma$.\\
To discuss this issue the limiting case of a so-called~\textit{interface problem} is considered, where the entire fluid domain $\Omega$ is separated by the interface~$\Gamma$ into two disjoint regions. The setting is equivalent to a small finite volume $\omega\subset\Omega$ which is entirely cut by the interface $\gamma\subset\Gamma$ into two subdomains $\omega^+\subset\omega$ and~$\omega^-\subset\omega$, as illustrated in Fig.~\ref{fig:domain_definition}.
The fluid inside~$\omega$ is described by the Navier-Stokes equations~\eqref{eq:NSE}, where the velocity field $\vec{u}^+$ and $\vec{u}^-$ belongs to the associated subdomain $\omega^+$ and $\omega^-$, respectively. The linear momentum balance~\eqref{eq:IGL} can now be formulated for $\omega^+$ and $\omega^-$ separately. Adding these results in a momentum balance for the union $\omega = \omega^+ \cup \omega^-$ results in 
\begin{equation}
  \label{eqn:cutmb}
	\int_{\omega} \left(\rho_\f\, \frac{\mathrm{d}\vec{u}}{\mathrm{d}t} - \vec{f}_V\right) \d V = \int_{\partial\omega} \mat{\sigma}\cdot\vec{n} \d S + \int_{\gamma} (\mat{\sigma}^+ \!- \mat{\sigma}^-)\cdot\vec{n} \d S \, ,
\end{equation}
where $\mat{\sigma}^+$ and $\mat{\sigma}^-$ are the stress tensors in $\omega^+$ and $\omega^-$, respectively. The second term on the right-hand side of Eq.~\eqref{eqn:cutmb} is referred to as \textit{jump term}. The associated jump in the stress vectors, i.e. $(\mat{\sigma}^+ \!- \mat{\sigma}^-)\cdot\vec{n}$, has the unit of a surface-specific force $\vec{f}_S$ related to the interface $\gamma$ \cite{Peskin1993,Li2006,Layton2008}. Using this force, the jump term can be expressed as
\begin{equation}
  \label{eqn:jump_surface_force}
   \int_{\gamma} \vec{f}_S \d S = \int_{\gamma}  \mat{\sigma}^+\cdot\vec{n} \:\text{d} S - \int_{\gamma}  \mat{\sigma}^-\cdot\vec{n} \d S \, ,
\end{equation}
which illustrates, that $\vec{f}_S$ is the resulting fluid load caused by the hydrodynamic stresses $\mat{\sigma}^+$ and $\mat{\sigma}^-$ acting on $\gamma$ from both sides, i.e. from $\omega^+$ and $\omega^-$, respectively. Only in cases where the stresses across $\gamma$ are discontinuous, the force $\vec{f}_S$ does not vanish. \\
As described in the previous section~\ref{subsec:coupling_condition}, the force $\vec{f}_S$ can also be interpreted as a coupling force introduced into the Navier-Stokes equations~\eqref{eq:NSE} and the Cosserat rod equations~\eqref{eqn:Cosserat} to impose the kinematic and dynamic coupling condition. In this context, the coupling force is defined locally at each point on $\vec{x}\in\Gamma$, opposed to its integral formulation according to Eq.~\eqref{eqn:jump_surface_force}. However, if the control volume $\omega$ shown in Fig.~\ref{fig:domain_definition} is decreased to an infinitesimally small size, Eq.~\eqref{eqn:jump_surface_force} can be transferred to a local relation between the coupling force and the hydrodynamic stresses, so that $\vec{f}_S=(\mat{\sigma}^+ \!- \mat{\sigma}^-)\cdot\vec{n}\;\;\forall \vec{x}\in\Gamma$. At the interface edge $\partial\Gamma$ the local force $\vec{f}_S$ vanishes, since $\mat{\sigma}^+$ and $\mat{\sigma}^-$ coincide in the free flow region without $\Gamma$.\\
The analysis of the cut volume $\omega$ via the momentum balance~\eqref{eq:IGL} revealed that the hydrodynamic stresses $\mat{\sigma}$ are discontinuous at $\gamma$, i.e. $(\mat{\sigma}^+ \!- \mat{\sigma}^-)\cdot\vec{n}\ne\vec{0}$, which implies that the velocity field does not need to be differentiable at $\Gamma$.

\subsubsection{Coupling via distributive sources}
\label{sec:dist source}
In the previous sections, the coupling force $\vec{f}_S=(\mat{\sigma}^+ \!- \mat{\sigma}^-)\cdot\vec{n}$ was derived as a surface-specific force acting on the fluid-structure interface $\Gamma$ with the associated coupling term appearing as an additional surface integral in the momentum balance~\eqref{eqn:cutmb} of the fluid. The IBM coupling strategy used here is based on the idea of converting this surface integral into a volume integral proposed by the present authors in~\cite{Tschisgale2017}. For this purpose, the surface-specific coupling force $\vec{f}_S$ is transformed into a mass-specific coupling force $\vec{f}_\IBM$. As a result, $\vec{f}_\IBM$ can be introduced directly into the differential momentum balance~\eqref{eq:IGL} as a regular volume force, such as gravitational acceleration. This simplifies the numerical treatment of the Navier-Stokes equations~\eqref{eq:NSE}, since a special handling of surface-specific quantities is not necessary.\\
The transformation of the integral coupling term is realized by using common techniques of distribution theory. In this context, a properly defined delta function $\delta_V$ allows replacing the surface-specific force $\textbf{f}_S$ by its volumetric complement $\textbf{f}_\IBM$~\cite{Onural2006,Farassat1996}, in such a way that the same momentum is transferred to the fluid, i.e.
\begin{equation}
	\label{eqn:surface_to_volume_integrals}
	\int_{\gamma} \vec{f}_S \d S = \int_{\omega} \delta_V \, \vec{f}_S \d V  = \int_{\lambda} \rho_\f\,\vec{f}_\IBM \: \text{d}V  \, ,
\end{equation}
where $\lambda\subset\Lambda$ is the compact, volumetric support of the delta function $\delta_V$ enveloping the interface $\gamma\subset\Gamma$ in the cut volume $\omega\subset\Omega$ (Fig.~\ref{fig:domain_definition}). On a larger scale, this support constitutes a thin ``coating'' layer $\Lambda$ enclosing entirely the interface $\Gamma$. For the exact continuous problem the thickness of $\Lambda$, $d_\Lambda$, is infinitesimally small, i.e. $d_\Lambda\rightarrow 0$. Thus, $\vec{f}_\IBM$ applies a finite amount of ``force'' to the fluid in an arbitrarily thin layer $\Lambda$, which indicates the \textit{distributive nature} of the coupling force. This perspective constitutes the basis for the discrete formulation of the FSI problem via an immersed boundary method, described in section~\ref{cha:cfsi} below.\\
According to the principle of~\textit{actio et reactio}, the distributive force $\vec{f}_\IBM$, introduced into the momentum balance of the fluid, must also appear in the equations of motion of the Cosserat rod, with opposite sign. So far, the forces $\fext_\IBM$ and moments $\mext_\IBM$ acting on the rod are defined via~$\vec{f}_S$, according to Eq.~\eqref{eqn:fluid_loads_3D}. Considering the zero-thickness approximation with $\Gamma\cap\Upsilon = \Upsilon$, both are given~by
\begin{equation}
  \label{eqn:fluid_force_local}
	\fext_\IBM(Z) = -\int_{\Upsilon} \vec{f}_S \d Y \,, \qquad 
  \mext_\IBM(Z) = -\int_{\Upsilon} \vec{\xi} \times \vec{f}_S \d Y \, ,
\end{equation}
with $\vec{f}_S=(\mat{\sigma}^+ \!- \mat{\sigma}^-)\cdot\vec{n}$. According to transformation~\eqref{eqn:surface_to_volume_integrals}, the coupling force $\vec{f}_\IBM$ emerges from a surface integration of $\vec{f}_S$. Instead, Eq.~\eqref{eqn:fluid_force_local} provides a line integral of $\vec{f}_S$ over $\Upsilon(Z)$. To connect $\fext_\IBM(Z)$ and $\mext_\IBM(Z)$ with the distributive force $\vec{f}_\IBM$, the former must be integrated along the arc length $Z$, i.e.
\begin{subequations}
	\begin{gather}
		\label{eqn:fluid_force_global}
		\int_{\zeta} \fext_\IBM \d Z = -\int_{\Gamma} \vec{f}_S \d S = -\int_{\Lambda} \rho_\f \: \vec{f}_\IBM \: \text{d} V \\
		\label{eqn:fluid_moment_global}
		\int_{\zeta} \mext_\IBM \d Z = -\int_{\Gamma} \vec{\xi} \times \vec{f}_S \d S = -\int_{\Lambda} \vec{\xi} \times (\rho_\f \: \vec{f}_\IBM) \: \text{d} V \,,
	\end{gather}
\end{subequations}
taking advantage of the fact that the interface $\Gamma$ equals the union of the cross-sections $\Upsilon(Z)$ (Fig.~\ref{fig:Cosserat_setup_zero}), i.e. $\Gamma = \bigcup_{Z\in\zeta} \!\Upsilon(Z)$. At a later stage, this relation is of crucial importance for the spatial discretization of the Cosserat rod by a finite set of structural elements $e$. In this context, the average hydrodynamic loads acting on the element interface $\Gamma_e$ are given by $\fext_{\IBM_e}\!=\int_{\zeta_e} \!\fext_\IBM\! \d Z / \Delta Z$ and $\mext_{\IBM_e}\!=\int_{\zeta_e} \!\mext_\IBM\! \d Z / \Delta Z$, with $\zeta_e\subset\zeta$ and $\Delta Z=\int_{\zeta_e}\!\! \d Z$.

\section{Numerical discretization of the partitioned problems}
\label{cha:numerics}

\subsection{Navier Stokes equations}
\label{sec:NSE_numerics}

\subsubsection{Temporal and spatial discretization}
The method proposed here was implemented in the in-house code PRIME (Phase-Resolving sIMulation Environment)~\cite{Kempe2012,Tschisgale2017}. The time integration of the Navier-Stokes equations~\eqref{eq:NSE} is accomplished by a special variant of the pressure projection method, used to impose the incompressibility constraint~\eqref{eq:KGL}. It combines an explicit three-step third-order low-storage Runge-Kutta scheme for the convective term and a second-order implicit Crank-Nicolson scheme for the viscous term in each Runge-Kutta sub-step. This variant conserves an overall second order accuracy in time for both, the pressure as well as the velocity~\cite{Brown2001}. Numerical stability of the time scheme is achieved for Courant numbers $\CFL<\sqrt{3}$~\cite{Rai1991}.
In each Runge-Kutta sub-step $r=1,2,3$ the following equations are solved:
\begin{subequations}
\label{eqn:fluid_solver}
\begin{gather}
	\label{eqn:RK_expl}
  \frac{ \tilde{\vec{u}} - \vec{u}^{r-1}}{\Delta t} =  2\alpha_{r}\: \nu_\f \:\nabla^{2}\vec{u}^{\:r-1}
  - 2\alpha_{r} \: \nabla \!\left(\frac{p}{\rho_\f}^{\! r-1}\right) - \gamma_{r} \:\nabla \cdot \left(\vec{u}\otimes\vec{u} \right) ^{\:r-1}
  - \zeta_{r} \: \nabla \cdot \left(\vec{u}\otimes\vec{u} \right) ^{\:r-2} + \vec{f}_V \\
	\label{eqn:helm+f}
	\nabla^{2}\vec{u}^{*} - \frac{\vec{u}^{*}}{\alpha_{r} \:\nu_\f \:\Delta t} = \nabla^{2}\vec{u}^{r-1} 
	- \frac{ \tilde{\vec{u}} + 2\alpha_r\,\Delta t\:\bar{\vec{f}}_\IBM }{\alpha_{r} \:\nu_\f \:\Delta t} \\
	\label{eqn:ibm_proj_1}
	\nabla^{2} \phi^{r} = \nabla \cdot \vec{u}^{*} \\
	\label{eqn:ibm_proj_2}
	\vec{u}^{r} = \vec{u}^{*} - \nabla \phi^{r} \\
	\label{eqn:ibm_proj_3}
	\frac{p}{\rho_\f}^{\! r} = \frac{p}{\rho_\f}^{\! r-1} + \frac{\phi^{r}}{ 2\:\alpha_{r}\:\Delta t} - \frac{\nu_\f}{2} \:\nabla^{2} \phi^{r} \, ,
\end{gather} 
\end{subequations}
with $\Delta{t}$ the time step and the values of the coefficients $\alpha_r$, $\gamma_r$ and $\zeta_r$ from~\cite{Rai1991}. The force $\bar{\vec{f}}_\IBM$ in Eq.~\eqref{eqn:helm+f} is used at a later stage for the fluid-structure coupling, described in section~\ref{cha:cfsi}.\\ 
The spatial discretization of the system~\eqref{eqn:fluid_solver} is performed by a second-order finite-volume scheme of Ham~\cite{Ham2002} on a Cartesian staggered grid. The discretization scheme fully conserves mass, momentum and energy even on non-uniform grids and avoids checkerboard oscillations of the pressure~\cite{Kempe2011_diss}. In the present work, a rectangular computational domain $\Omega = [0;L_x] \times [0;L_y] \times [0;L_z]$ is used, where $L_x$, $L_y$ and $L_z$ denote the extend of the domain in $x$-, $y$- and $z$-direction, respectively. The usual boundary conditions for the velocity field are available. For the pressure correction field $\phi$ a zero gradient condition \cite{Gresho1987} is used for all types of boundary conditions except for periodic boundaries.

\subsubsection{Large eddy simulation and subgrid-scale model}
The present numerical approach is designed to provide time-dependent high-resolution data, which are utilized to study the dynamic behavior of the coupled system and the role of turbulent structures, also for higher turbulence intensities of the fluid. Under the latter conditions, the direct numerical simulation (DNS) of the Navier-Stokes equations~\eqref{eq:NSE} is technically not feasible with the present discretization technique since the required grid resolution can not be achieved. In such cases, the large eddy simulation (LES) approach is employed here, using the Smagorinsky model~\cite{Smagorinsky1963} to model subgrid-scale stresses. In the present work it is assumed, that the grid spacing is fine enough to capture both the flow near the walls and at the fluid-structure interface $\Gamma$, so that no additional modeling is required, e.g. by a wall function. Non-physical values of the eddy viscosity in the region close to walls are reduced by a Van Driest damping function~\cite{VanDriest1956}. 

\subsection{Cosserat rod equations}
\label{sec:Cosserat_numerics}
\subsubsection{Parametrization of finite rotations}

The Cosserat rod equations~\eqref{eqn:Cosserat_angular} are constituted by a linear and an angular equation of motion. The former describes the motion of the center line position $\vec{c}\in\mathbb{R}^3$. The angular equation of motion is formulated in terms of the rotation matrix $\mat{R} \in SO(3)$ additionally subjected to the properties of the rotation group $SO(3)$, i.e. the orthogonality constraint $\mat{R}\cdot\mat{R}^\top\!=\mat{R}^\top\cdot\mat{R}=\mathbb{I}$. This constraint is taken into account when parameterizing $\mat{R}$. State of the art is to describe rotations via quaternions $\vec{q} \in \mathbb{S}^3$ with the set of unit quaternions $\mathbb{S}^3$ \cite{Kuipers1999,Lang2011}. Among other advantages, these avoid the {\it gimbal lock} effect or singularities~\cite{Kuipers1999}. Lang~\textit{et al.}~\cite{Lang2011} showed that the Cosserat rod equations~\eqref{eqn:Cosserat} can be reformulated by standard index reduction techniques as an equivalent system
\begin{subequations}
	\label{eqn:Cosserat_quat}
	\begin{align}
		\label{eqn:Cosserat_quat_linear}
			\ddot{\vec{c}} &= \,\frac{1}{\rho_\s A}\,\left\{\,\left(\vec{q}\qm\fint\tra\!\qm\bar{\vec{q}}\right)' + \fext\,\right\} \\
		\label{eqn:Cosserat_quat_angular}
			\ddot{\vec{q}} &= \frac{2}{\rho_\s}\,\mat{\mathcal{M}}\cdot\left\{\,4\rho_\s \,\dot{\vec{q}}\qm\mat{\mathcal{I}}\tra\!\cdot\!\left(\dot{\bar{\vec{q}}}\qm\vec{q}\right) + \vec{c}'\!\qm\vec{q}\qm\fint\tra + \left(\vec{q}\qm\mint\tra\right)' + \vec{q}'\!\qm\mint\tra + \mext\qm\vec{q} \,\right\} - \|\dot{\vec{q}}\|^2\vec{q} \, ,
	\end{align}
\end{subequations}
with the quaternion matrix of inertia $\mat{\mathcal{I}}\tra\!=0\oplus\mat{I}\tra$ and the ``inverse'' quaternion matrix of inertia $\mat{\mathcal{M}} = \frac{1}{4}\,\mat{\mathcal{Q}}\cdot\mat{\mathcal{I}}^{-1}\tra\cdot\mat{\mathcal{Q}}^\top$. Here, the matrix $\mat{\mathcal{Q}}$ allows to express a multiplication of $\vec{p},\vec{q}\in\mathbb{S}^3$ as a matrix-vector product, i.e. $\vec{p}\,\qm\,\vec{q}=\mat{\mathcal{Q}}(\vec{p})\cdot\vec{q}$~\cite{Lang2011}. Vectors indicated with index zero, e.g. $\vec{v}\tra \!\in \mathbb{R}^3$, are given in the local co-rotated Lagrangian frame of a cross section and quantities without index, such as $\vec{v} \in \mathbb{R}^3$, in the global Eulerian frame. According to Eq.~\eqref{eqn:internal_loads} internal forces $\fint\tra$ and internal moments $\mint\tra$ are formulated in terms of the strain vector $\vec{\varepsilon}\tra$ and the curvature vector $\vec{\kappa}\tra$, respectively. Their equivalent quaternionic forms are given by
\begin{equation}
	\label{eqn:strains_quat}
	\vec{\varepsilon}\tra\!= \bar{\vec{q}}\qm\vec{c}'\qm\vec{q} \qquad \text{and} \qquad \vec{\kappa}\tra\!=2\,\bar{\vec{q}}\qm\vec{q}' \, .
\end{equation}

\subsubsection{Temporal and spatial discretization}
\label{sec:discretization_rod}
\begin{figure}[!tb]
 \centering
 \includegraphics[trim=0 0 0 0, clip, scale=1]{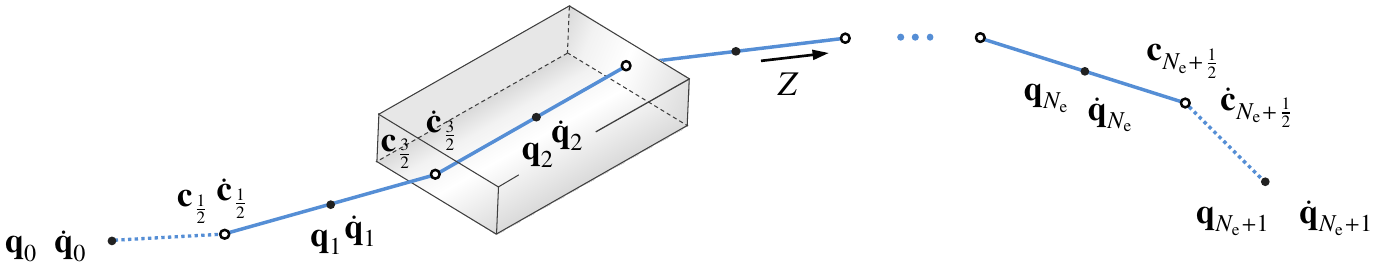}
 \caption{\label{fig: discretization rod}Spatial discretization of the rod along the arc length $Z$ by $N_\mathrm{e}$ elements of equal geometrical and material properties. Shown are discrete centroid positions $\vec{c}_{e-\frac{1}{2}}$ and velocities $\dot{\vec{c}}_{e-\frac{1}{2}}$, as well as the rotational degree of freedom represented by quaternions $\vec{q}_e$ and $\dot{\vec{q}}_e$, with $e=1,...,N_\mathrm{e}$. The dashed lines indicate the connection to ghost-quaternions $\vec{q}_0$ and ${\vec{q}}_{N_\mathrm{e}+1}$ and ghost-velocities $\dot{\vec{q}}_0$ and $\dot{\vec{q}}_{N_\mathrm{e}+1}$ required to impose the boundary conditions at both ends.}
\end{figure}

Following recommendations of \text{Lang~\textit{et al.}~\cite{Lang2011}}, the rod equations~\eqref{eqn:Cosserat_quat} are discretized by the finite difference methods (FDM) using an equidistant staggered grid to achieve a second order accuracy in space. As shown in Fig.\ref{fig: discretization rod}, the centroids of the cross sections are located at the edges of an element $e$ which are denoted by a half-index, i.e. $\vec{c}_{e-\frac{1}{2}}$ and $\vec{c}_{e+\frac{1}{2}}$ with $e=1,...,N_\mathrm{e}$.	
The spatial discretization of the Cosserat equations via FDM results in a system of first-order ordinary differential equations (ODE) of the form
\begin{equation}
	\label{eqn:ode}
	\dot{\vec{\mathfrak{u}}}=\vec{rhs}(\vec{\mathfrak{u}},t), \quad
	\vec{\mathfrak{u}}=\left(\vec{c}_\frac{1}{2},\dot{\vec{c}}_\frac{1}{2},\;\dot{\vec{q}}_1,\vec{q}_1,\;\vec{c}_\frac{3}{2},\dot{\vec{c}}_\frac{3}{2},\,\,...\,\,,\,\dot{\vec{q}}_{N_\mathrm{e}},\vec{q}_{N_\mathrm{e}},\;\vec{c}_{N_\mathrm{e}+\frac{1}{2}},\dot{\vec{c}}_{N_\mathrm{e}+\frac{1}{2}}\right)^\top .
\end{equation}
The time integration of system~\eqref{eqn:ode} can be done by an appropriate ODE-solver, adapted to the numerical properties of the ODE. Following recommendations of Lang et~al.~\cite{Lang2011} the non-commercial solver RADAU5~\cite{Hairer1999,RadauDopri} is employed here. Successfully solving the differential equations~\eqref{eqn:Cosserat_quat} does not necessarily impose the constraint of unit length, i.e. $\|\vec{q}\|=1$, required to describe rotations in space. As described in~\cite{Lang2011}, the quaternion $\vec{q}$ drifts quadratically from this constraint which, however, can be counteracted by the projection
\begin{equation}
	\label{eqn:projection}
	\vec{q} \leftarrow \vec{q}/\|\vec{q}\| \qquad \text{and} \qquad \dot{\vec{q}} \leftarrow \dot{\vec{q}} - (\vec{q}\!\cdot\!\dot{\vec{q}})\,\vec{q} \, ,
\end{equation}
applied after each time integration step for the entire set of quaternions $\vec{q}^{n+1}_e$, $e=1,...,N_\mathrm{e}$.

\section{Semi-implicit direct forcing IBM}
\label{cha:cfsi}

\subsection{Temporal coupling using direct forcing}
\label{sec:direct forcing}

\subsubsection{The direct forcing approach}
Different variants of the immersed boundary method can be distinguished by the way in which the coupling force $\vec{f}_\IBM$ in Eq.~\eqref{eq:IGL} is computed. As already mentioned, a spatially continuous force $\vec{f}_\IBM$, acting in an infinitesimally thin layer $\Lambda$ (Fig.~\ref{fig:domain_definition}), is employed in the present work to impose the kinematic and dynamic coupling condition. In the IBM framework this approach is usually denoted as continuous forcing~\cite{Mittal2005}. From a numerical point of view two aspects are decisive. First, the local force has to be evaluated in a time discrete manner to realize a coupling of fluid and structures, called temporal coupling here. Second, a suitable approach is required for a spatial transfer of information between the fluid and the structure, each discretized in a different manner. While the temporal coupling is described in this section, the next section focuses on the spatial coupling. Finally, both approaches for temporal and spatial coupling are combined in section~\ref{sec:coupling scheme}, thus providing the complete coupling algorithm.\\
In the present work, the temporal coupling is realized by the direct forcing approach~\cite{Mohd-yusof1997,Fadlun2000,Uhlmann2005}. Its basic idea is to incorporate the no-slip condition on $\Gamma$ at a time discrete level to determine the coupling force within a certain time interval $t \in [t^{n},t^{n+1}]$. According to the momentum balance~\eqref{eq:IGL} the coupling force at $\vec{x}\in\Gamma$ is
\begin{equation}
	\label{eqn:f_ibmdf}
	\vec{f}_\IBM = \frac{\partial{\vec{u}}}{\partial{t}} - \vec{rhs} \, ,
\end{equation}
where the right-hand side $\vec{rhs}$ includes the convective, pressure and viscous terms~\cite{Mohd-yusof1997}. The coupling force can then be obtained by integrating Eq.~\eqref{eqn:f_ibmdf} with an arbitrary time-stepping scheme over the time interval $t \in [t^{n},t^{n+1}]$
\begin{equation}
	\label{eqn:f_ibm_int}
	\int_{t^{n}}^{t^{n+1}} \vec{f}_{\IBM} \d t = \int_{t^{n}}^{t^{n+1}} \left(\frac{\partial{\vec{u}}}{\partial{t}} - \vec{rhs}\right) \d t 
	= \vec{u}^{n+1} \!- \vec{u}^{n} - \int_{\:t^{n}}^{\:t^{n+1}} \vec{rhs}\: \text{d} t \, ,
\end{equation}
with $\vec{u}^n$ and $\vec{u}^{n+1}$ being the fluid velocities at time level $t^{n}$ and $t^{n+1}$, respectively. By incorporating the no-slip condition \eqref{eqn:no-slip_condition}, $\vec{u}^{n+1}$ is replaced by the local desired velocity $\vec{u}_\Gamma^{n+1}$ of the interface~$\Gamma$, yielding
\begin{equation}
	\label{eqn:f_ibm_int_3}
	\int_{t^{n}}^{t^{n+1}} \vec{f}_{\IBM} \d t = \vec{u}_\Gamma^{n+1} \!- \vec{u}^{n}-  \int_{t^{n}}^{t^{n+1}} \vec{rhs} \d t  \qquad \forall\; \vec{x} \in \Lambda \, ,
\end{equation}
while $\vec{f}_{\IBM}$ vanishes at locations $\vec{x} \notin \Lambda$. In the literature, this equation is usually converted into
\begin{equation}
	\label{eqn:f_ibm_impl2}
	\bar{\vec{f}}_{\IBM} = \frac{1}{\Delta t} \: \int_{t^{n}}^{t^{n+1}} \vec{f}_{\IBM} \d t= \frac{ \vec{u}_\Gamma^{n+1}  - \tilde{\vec{u}}}{\Delta t} \, ,
\end{equation}
where $\bar{\vec{f}}_{\IBM}$ is the average coupling force applied over the time interval $t \in [t^{n},t^{n+1}]$ and $\tilde{\vec{u}}$ is a shorthand for
\begin{equation}
	\label{eqn:prelim_vel}
	\tilde{\vec{u}} = \vec{u}^{n} + \int_{t^{n}}^{t^{n+1}} \vec{rhs} \d t \, ,
\end{equation}
which is the preliminary velocity of the fluid obtained without accounting for the effect of the immersed boundary.

\subsubsection{Modified equation of motion}
As described in section~\ref{subsec:coupling_condition}, the fluid-structure coupling is accomplished by two conditions, the kinematic coupling condition~\eqref{eqn:no-slip_condition} and the dynamic coupling condition~\eqref{eqn:dynamic_condition}. Using the direct forcing method, the former is incorporated directly into the coupling force $\vec{f}_{\IBM}$ imposing the no-slip condition in the fluid field, i.e. $\vec{u}=\vec{u}_\Gamma$ at $\vec{x}\in\Gamma$. In accordance with the dynamic coupling condition and the principle of \textit{actio et reactio}, the coupling force $\vec{f}_{\IBM}$ also appears in the equation of motion of the immersed boundary $\Gamma$ with opposite sign. As a result, the motion of $\Gamma$ and the coupling force~\eqref{eqn:f_ibm_impl2} exhibit an implicit dependency, since $\vec{f}_{\IBM}$ is a function of $\vec{u}_\Gamma^{n+1}$ at the new time level $t^{n+1}$. This becomes clearer when considering a general motion of $\Gamma$ described by the differential equation
\begin{equation}
  \label{eqn:structure_motion_dgl}
	\dot{\vec{u}}_\Gamma = \vec{rhs}_\Gamma(\vec{u}_\Gamma,t) - \pi_\rho \vec{f}_{\IBM} \qquad \forall \vec{x}\in\Gamma,
\end{equation}
with the right-hand side $\vec{rhs}_\Gamma$ describing the unconstrained motion of $\Gamma$, coupled to the fluid via $\pi_\rho\vec{f}_{\IBM}$. Here, $\pi_\rho \geq 0$, specifies the ratio of inertia between the fluid and the immersed structure represented by the boundary $\Gamma$. After integrating in time and using Eq.~\eqref{eqn:f_ibm_impl2} the discrete motion is given by
\begin{equation}
	\label{eqn:structure_motion_discrete}
	\vec{u}^{n+1}_\Gamma = \vec{u}_\Gamma^{n} + \int_{t^{n}}^{t^{n+1}} \vec{rhs}_\Gamma(\vec{u}_\Gamma,t) \d t \;-\; \Delta t\, \pi_\rho \bar{\vec{f}}_{\IBM}(\vec{u}_\Gamma^{n+1}\!,\tilde{\vec{u}}) 
	\, .
\end{equation}
Two strategies are now possible to solve this equation. The first is based on the observation that $\bar{\vec{f}}_{\IBM}$ depends on the difference $\vec{u}^{n+1}_\Gamma - \tilde{\vec{u}}$. This suggests to bring the contribution of $\bar{\vec{f}}_{\IBM}$ depending on $\vec{u}^{n+1}_\Gamma$ to the left-hand side, resulting in a kind of added mass effect. This strategy was employed for FSIs with one-dimensional filaments by Xu~\textit{et al.}~\cite{Xu2018} and earlier by Tschisgale~\mbox{\textit{et al.}~\cite{Tschisgale2016,Tschisgale2017}} for rigid bodies. However, for Cosserat rods or similar three-dimensional structures the additional mass term resulting from $\bar{\vec{f}}_{\IBM}$ may become very complex and requires considerable manipulation of the time-discrete equation of motion. The second strategy proposed here avoids this complication. When using particular libraries for an integration of the structure equation, the ODE integrators usually only provide an interface to the continuous version of the differential equation, e.g. $\dot{\vec{u}}_\Gamma = \vec{rhs}_\Gamma(\vec{u}_\Gamma,t)$, and not an already discrete version including $\vec{u}_\Gamma^{n+1}$ in the coupling force~$\bar{\vec{f}}_{\IBM}$. The user simply has to provide a continuous function of $\vec{rhs}_\Gamma(\vec{u}_\Gamma,t)$ to the solver while using it as a black-box without specific technical knowledge of the discretization scheme employed. To realize the direct forcing approach with a standard black-box ODE solver, a continuous version of Eq.~\eqref{eqn:structure_motion_discrete} is required. While $\vec{rhs}_\Gamma(\vec{u}_\Gamma,t)$ is already known, the coupling force needs to be reformulated as an expression $\vec{f}_{\IBM}(\vec{u}_\Gamma,t)$ that continuously depends on time and on the interface velocity. Doing so, Eq.~\eqref{eqn:structure_motion_discrete} can be expressed as a modified equation of motion of~$\Gamma$
\begin{equation}
\label{eqn:modified_equation}
	\dot{\vec{u}}_\Gamma = \vec{rhs}_\Gamma(\vec{u}_\Gamma,t) - \pi_\rho \vec{f}_{\IBM}(\vec{u}_\Gamma,t) \,=\, {\vec{rhs}}_{\Gamma,\mathrm{mod}}(\vec{u}_\Gamma,t) \, ,
\end{equation}
which can simply be passed to an arbitrary implicit ODE solver without need of knowledge about the time discretization technique. In the present work, the continuous variant of the coupling force is obtained by considering a linear behavior of the interface velocity within the given time interval $t \in [t^{n},t^{n+1}]$, i.e. 
\begin{equation}
  \label{eqn:ugamma_lin}
	\vec{u}_{\Gamma,\mathrm{lin}}(t) = \left(\vec{u}^{n+1}_\Gamma - \vec{u}^{n}_\Gamma\right)\,\frac{t-t^n}{\Delta t} + \vec{u}^{n}_\Gamma \, ,
\end{equation}
which can be rearranged into
\begin{equation}
	\vec{u}^{n+1}_\Gamma = \left(\vec{u}_{\Gamma,\mathrm{lin}} - \vec{u}^{n}_\Gamma\right)\,\frac{\Delta t}{t-t^n} + \vec{u}^{n}_\Gamma \, ,
\end{equation}
such that the interface velocity at the new time level $t^{n+1}$ is provided as a continuous function $\vec{u}^{n+1}_\Gamma = \vec{u}^{n+1}_\Gamma\!\left(\vec{u}_\Gamma,t\right)$. Using this formulation in combination with the direct forcing approach~\eqref{eqn:f_ibm_impl2}, the coupling force in Eq.~\eqref{eqn:modified_equation} can be approximated by
\begin{equation}
	\label{eqn:lin_IBM_force}
	\vec{f}_{\IBM}(\vec{u}_\Gamma,t) \approx \vec{f}_{\IBM,\text{lin}}(\vec{u}_\Gamma,t) = \frac{\vec{u}_\Gamma - \vec{u}^{n}_\Gamma}{t-t^n} + \frac{\vec{u}^{n}_\Gamma - \tilde{\vec{u}}}{\Delta t} \, .
\end{equation}
As a crosscheck, the time integration of $\vec{f}_{\IBM,\text{lin}}$ for a linear slope of $\vec{u}_\Gamma(t) = \vec{u}_{\Gamma,\text{lin}}(t)$ yields
\begin{equation}
	\frac{1}{\Delta t}\int^{t^{n+1}}_{t^n} \vec{f}_{\IBM,\text{lin}}(\vec{u}_{\Gamma,\text{lin}},t) \d t \; = \; \bar{\vec{f}}_\IBM = \frac{\vec{u}^{n+1}_\Gamma - \tilde{\vec{u}}}{\Delta t}
\end{equation}
and, thus, reproduces the common direct forcing according to Eq.~\eqref{eqn:f_ibm_impl2}. Since the preliminary velocity $\tilde{\vec{u}}$ in $\vec{f}_{\IBM,\mathrm{lin}}(\vec{u}_\Gamma,t)$ is computed for a particular time interval $[t^{n},t^{n+1}]$, the coupling force and the corresponding modified right-hand side of Eq.~\eqref{eqn:modified_equation} is valid only for this specific time interval as well.\\
In general, the modified equations of motion~\eqref{eqn:modified_equation} can be formulated for any kind of immersed boundary, ranging from rigid bodies to deformable structures, by adding the continuous version of the coupling force $\vec{f}_{\IBM,\text{lin}}(\vec{u}_\Gamma,t)$ to the ODE describing the decoupled motion of $\Gamma$.

\subsubsection{Application to Cosserat rod equations.}
In the present context, the motion of the immersed boundary $\Gamma$ is described by the Cosserat rod equations~\eqref{eqn:Cosserat} that contain external fluid forces $\fext_\IBM$ and external fluid moments $\mext_\IBM$, both related to the coupling force $\vec{f}_{\IBM}$ via Eqs.~\eqref{eqn:fluid_force_global} and~\eqref{eqn:fluid_moment_global}, respectively.
To apply the ``black-box'' technique via $\vec{f}_{\IBM,\text{lin}}$ just mentioned, the dependency of $\vec{u}_\Gamma$ in Eq.~\eqref{eqn:lin_IBM_force} must be expressed in terms of velocity quantities provided by the Cosserat rod, i.e. the linear velocity of the center line $\dot{\vec{c}}$ and the angular velocities represented by the quaternionic velocity $\dot{\vec{q}}$.
According to the no-slip condition~\eqref{eqn:no-slip_condition2} the velocity at the interface $\Gamma$ is given by
\begin{subequations}
\begin{align}
	\label{eqn:ugamma_1}
  \vec{u}_\Gamma &= \dot{\vec{c}} + \vec{\omega} \times \vec{\xi} \\
	\label{eqn:ugamma_3}
	&= \dot{\vec{c}} + \left(2\,\dot{\vec{q}}\qm\bar{\vec{q}}\right) \times \left({\vec{q}}\qm\vec{\xi}_{\text{\tiny{0}}}\qm\bar{\vec{q}}\right) \, ,
\end{align}
\end{subequations}
where in the second variant~\eqref{eqn:ugamma_3} the angular velocity is expressed by means of quaternions via $\vec{\omega}=2\,\dot{\vec{q}}\,\qm\,\bar {\vec{q}}$. Furthermore, the vector $\vec{\xi}$ can be rotated backwards into the local frame, so that $\vec{\xi} = {\vec{q}}\,\qm\,\vec{\xi}_{\text{\tiny{0}}}\qm\,\bar{\vec{q}}$. Since $\vec{\xi}_{\text{\tiny{0}}}$ in the local reference frame is time-independent, the continuous coupling force~\eqref{eqn:lin_IBM_force} can be expressed as $\vec{f}_{\IBM,\text{lin}}(\dot{\vec{c}},\dot{\vec{q}},\vec{q},t)$. Using this force, the related external fluid forces $\fext_\IBM$~\eqref{eqn:fluid_force_global} and moments $\mext_\IBM$~\eqref{eqn:fluid_moment_global} can be approximated~by
\begin{subequations}
\label{eqn:coupling_terms_mod}
\begin{align}
\label{eqn:coupling_terms_mod1}
	\int_{\zeta} \fext_\IBM \d Z &\approx -\int_{\Lambda} \rho_\f \: \vec{f}_{\IBM,\text{lin}}(\dot{\vec{c}},\dot{\vec{q}},\vec{q},t) \: \text{d} V \\
	\label{eqn:coupling_terms_mod2}
  \int_{\zeta} \mext_\IBM \d Z &\approx -\int_{\Lambda} \vec{\xi} \times \bigl[\rho_\f \: \vec{f}_{\IBM,\text{lin}}(\dot{\vec{c}},\dot{\vec{q}},\vec{q},t)\bigl] \: \text{d} V
\end{align}
\end{subequations}
in the time interval $t \in [t^{n},t^{n+1}]$.
		
\subsubsection{Resulting coupling scheme}
\label{sec:resulting_scheme}
The modified direct forcing proposed in the previous section can be summarized by the following steps:
\begin{enumerate}
 \item Computation of the preliminary velocity $\tilde{\vec{u}}$ via Eq.~\eqref{eqn:prelim_vel} without accounting for any coupling to the immersed interface $\Gamma$.
 \item Communication of the preliminary velocity $\tilde{\vec{u}}(\vec{x})$ at $\vec{x}\in\Gamma$ to the structure solver.
 \item Computation of the interface velocity $\vec{u}^{n+1}_\Gamma$ at the new time level $t^{n+1}$ by solving the equation of motion~\eqref{eqn:modified_equation} implicitly, modified by the coupling force $\vec{f}_{\IBM,\text{lin}}$~\eqref{eqn:lin_IBM_force}.
 \item Determination of the coupling force $\bar{\vec{f}}_\IBM(\vec{x})$ at $\vec{x}\in\Gamma$ via Eq.~\eqref{eqn:f_ibm_impl2} and communication of $\bar{\vec{f}}_\IBM$ to the fluid solver.
 \item Solving the Navier-Stokes equations~\eqref{eq:NSE} coupled to the immersed boundary $\Gamma$ by $\bar{\vec{f}}_\IBM$.
\end{enumerate}
\begin{figure}[!htb]
	\centering
	\includegraphics[trim=0 0 0 0, clip, scale=1]{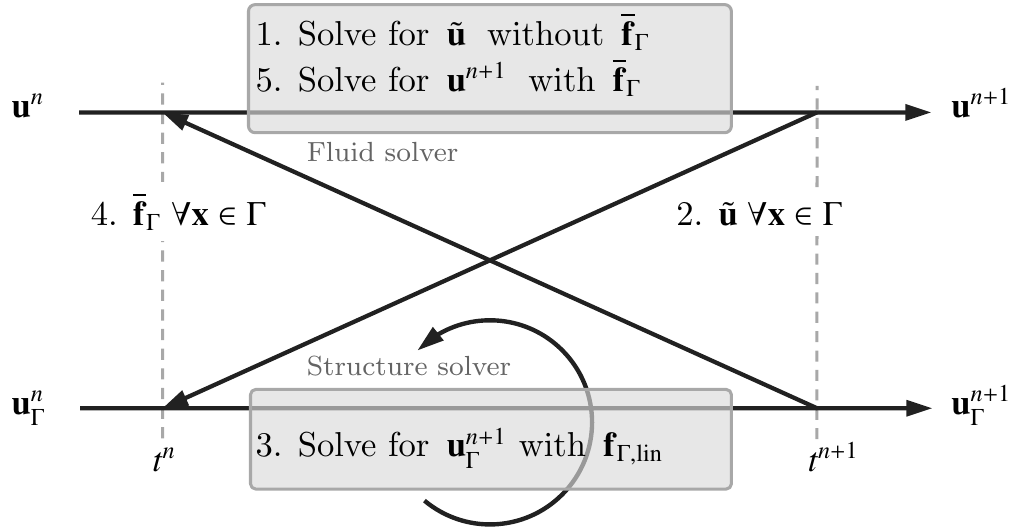}
  \caption{\label{fig:temp_coupling}Flowchart of the five steps to be performed for the temporal coupling of the fluid solver and the structure solver within one time step $t\in[t^n,t^{n+1}]$ by means of the direct forcing approach proposed here. The circular arrow illustrates the implicit nature of the solution procedure employed to solve for the structure motion. }
\end{figure}
The implicit treatment of the modified equation of motion of the structure in step 3 corresponds to a strong coupling of fluid and structure and achieves numerical stability for arbitrary immersed objects. It can be solved by an iterative procedure, e.g. a Newton method or an implicit Runge-Kutta scheme, such as RADAU5 employed in the present work.\\
As an alternative to an implicit treatment, the modified equations could be treated by an explicit integration scheme as well. When integrating the continuous coupling force $\vec{f}_{\IBM}(\vec{u}_\Gamma,t)$ according to Eq.~\eqref{eqn:lin_IBM_force} with an explicit Euler scheme, e.g., it simplifies to
\begin{equation}
	\frac{1}{\Delta t}\int^{t^{n+1}}_{t^n} \!\!\!\!\!\vec{f}_{\IBM,\text{lin}}(\vec{u}_{\Gamma,\text{lin}},t) \d t \;\;\; \approx \;\;\; \vec{f}_{\IBM,\text{lin}}(\vec{u}^n_{\Gamma,\text{lin}},t^n)  = \frac{\vec{u}^{n}_\Gamma - \tilde{\vec{u}}}{\Delta t} \, .
\end{equation}
Obviously, in contrast to the exact direct forcing~\eqref{eqn:f_ibm_impl2}, it is based on employing the interface velocity at the old time~$t^{n-1}$. In this case the forcing scheme is equal to the well-known variants of an explicit IBM, proposed, e.g., in~\cite{Uhlmann2005,Kempe2012,Breugem2012}. It is known that these variants become unstable, especially for lightweight immersed objects~\cite{Kempe2012}. In addition, as demonstrated in previous works~\cite{Tschisgale2016,Tschisgale2017}, replacing $\vec{u}^{n+1}_\Gamma$ with $\vec{u}^{n}_\Gamma$ may result in a numerically inconsistent coupling, where the numerical solution does not converge to the monolithic solution by a spatial and temporal refinement. The solution only coincides with the monolithic solution if the mass ratio between the structure mass and the mass of the surrounding fluid layer $\Lambda$ tends to infinity. However, for sufficiently large mass ratios the numerical error remains very small. For practical applications, the explicit direct forcing, based on $\vec{u}^{n}_\Gamma$, does not constitute any advantage over the present scheme in terms of implementation effort, numerical efficiency and accuracy.
In other words, the present scheme combines the stability properties of strong coupling schemes with the efficiency and ease of implementation of weak coupling schemes.\\
Since an analytical proof of the stability of the proposed coupling between the Navier-Stokes equations and Cosserat rod equations is out of reach, the assessment of the stability behavior is based on own simulation experiences. In all these, no stability issues were observed, including simulations with stiff and soft rods, high and moderate Reynolds numbers as well as FSI problems with strong added mass effects. The latter property is supported by previous FSI simulations with rigid bodies based on the above coupling strategy~\cite{Tschisgale2016, Tschisgale2017}. Even for rigid bodies of zero mass, where inertia effects are solely given by the added mass of the fluid, the proposed coupling works without any stability issues.\\
In addition to empirical test simulation, the stability of staggered coupling schemes is often verified by using a simple linear model problem, e.g. a mass-spring-damper model~\cite{Dettmer2013}. Here, a similar system is used for this purpose based on two coupled Dahlquist test equations, 
\begin{subequations}
  \label{eq:dahl}
\begin{align}
  \label{eq:dahl_1}
	\dot{u} &= \alpha u + f \\
	\label{eq:dahl_2}
  \dot{u}_\Gamma &= \alpha_\Gamma u_\Gamma - \pi_\rho f \\
	u &= u_\Gamma \quad \text{\small(coupling condition)}\, ,
\end{align}
\end{subequations}
where $\alpha < 0$, $\alpha_\Gamma < 0$ and $\pi_\rho \ge 0$, with Eq.~\eqref{eq:dahl_1} and~\eqref{eq:dahl_2} inspired by Eq.~\eqref{eqn:f_ibmdf} and~\eqref{eqn:structure_motion_dgl}, respectively. Within a time step $[t^{n},t^{n+1}]$ both equations are coupled by the above strategy involving the coupling force~\eqref{eqn:lin_IBM_force}. The implicit Euler scheme is employed for time integration which results in a discrete system of the form $(u^{n+1}, u_\Gamma^{n+1})^\top = \mat{A} \cdot (u^{n}, u_\Gamma^{n})^\top$, with the amplification matrix $\mat{A}$ depending on $\Delta t\alpha, \Delta t\alpha_\Gamma$ and $\pi_\rho$. It was checked that the spectral radius of $\mat{A}$ does not exceed one for any of these parameters. Hence, the proposed coupling strategy is unconditionally stable for the model system~\eqref{eq:dahl} discretized with an implicit Euler scheme.

\subsection{Spatial coupling via marker points}
\label{sec:marker}
\subsubsection{Lagrangian markers and volumes}
\label{sec:markers}
\begin{figure}[!tb]
	\centering
	\includegraphics[trim=0 0 0 0, clip, scale=1]{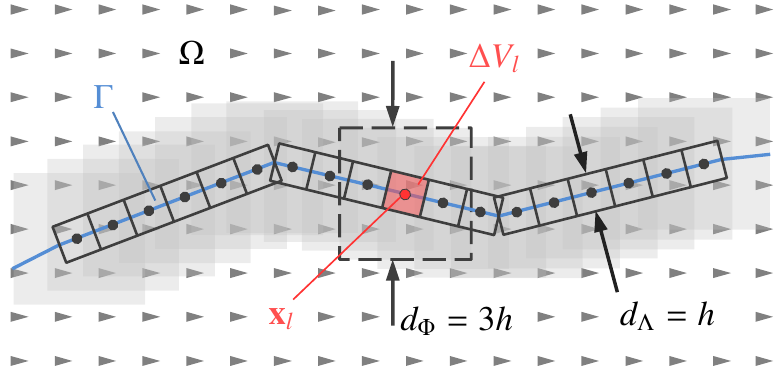}
  \caption{\label{fig:rod_discrete}Sketch of the spatial discretization employed by the immersed boundary method. The rod is represented by Lagrangian points $\vec{x}_l$. Each point is attributed a Lagrangian volume $\Delta V_l$ centered around $\vec{x}_l$. The union of all volumes constitutes a layer $\Lambda$ of width $d_\Lambda\!=h$. The connection between the Lagrangian points~$\vec{x}_l$ and the Eulerian grid $\vec{x}_{ijk}$ is realized by regularized delta functions, described below. The gray shaded area displays the cumulative spatial influence of all delta functions, one applied at each marker point. The dashed line represents the area of impact in the velocity field when the three-point delta function of Roma~\textit{et al.}~\cite{Roma1999} is employed. Note, that only the staggered grid for the velocity component $u$ is shown for simplicity. }
\end{figure}

In the framework of common IBMs the discrete elements of the structures do not coincide with the points of the Eulerian grid of the fluid. Hence, their coupling requires some technique to transfer information between both discrete representations. For this purpose, each zero-thickness rod is represented here by a set of discrete markers, so-called Lagrangian points, implementing this communication. The Lagrangian points $\vec{x}_l$, $l=1,...,N_l$, are evenly distributed over the fluid-structure interface $\Gamma$, as shown exemplarily in Fig.~\ref{fig:rod_fp}, below.
Furthermore, a two-dimensional sketch of the discretization of a rod embedded in the Eulerian background grid is shown in Fig.~\ref{fig:rod_discrete}.
The fluid-structure coupling is realized by a distributive coupling force $\vec{f}_\IBM$ acting in a small layer $\Lambda$ around the interface $\Gamma$. While in the continuous formulation of the coupling force~\eqref{eqn:surface_to_volume_integrals} the support of the corresponding delta function $\delta_V$ is infinitesimally small, i.e. $d_\Lambda \rightarrow 0$, in the discrete realization the thickness of the layer $\Lambda$ has to be equal to the step size of the Eulerian grid, i.e. $d_\Lambda=h$, as discussed in~\cite{Tschisgale2017}. Hence, at least one marker point controls a volume equal to the volume of a fluid cell. This means that each Lagrangian volume~$\Delta V_l$,~associated to a marker point, has to be chosen smaller or equal to the size of the Eulerian fluid cells, i.e.
\begin{equation}\label{eqn:Lagr_volume}
	\Delta V_l = \Delta S_l \: h \le h^3 \, ,
\end{equation}
where $\Delta S_l$ is the corresponding surface area attributed to a particular marker point. The volume of the entire layer, as the sum of all volumes $\Delta V_l$, fulfills the condition
$\sum_l \Delta V_l = S_\Gamma \: h$, with the surface area of the rod $S_\Gamma = \int_\Gamma \!\!\d S = \sum_l \Delta S_l $, yielding $N_l \geq S_\Gamma/h^2$.

\subsubsection{Regularized delta functions}
\label{sec:delta_functions}
The transfer of information between fluid and structure is performed via regularized delta functions $\delta_h$. As common for the present type of IBM the three-dimensional function $\delta_h$ is generated by a tensor product of three one-dimensional functions $\delta_h^\mathrm{1D}$, so that
\begin{equation} 
	\label{eqn:eq_deltah}
	\delta_h(\vec{r})\:=\:\delta_h^\mathrm{1D}(r_x) \:\: \delta_h^\mathrm{1D}(r_y) \:\: \delta_h^\mathrm{1D}(r_z)
\end{equation}
with the distance vector $\boldsymbol{\vec{r}}=(r_x,r_y,r_z)^\top\!$. Furthermore, $\delta_h^\mathrm{1D}(r_x)=\Phi(r)/h$ and $r=r_x/h$, etc. The continuous function $\Phi$ is constructed so as to fulfill certain properties, e.g. moment conditions~\cite{Peskin2002}, and several proposals have been made in the literature~\cite{Yang2009}. Here, the three-point version of Roma et~al.~\cite{Roma1999} 
\begin{equation}
 \label{eqn:dirac_roma}
	 \Phi_3(r) = 
   \begin{cases}
			\frac{1}{6}\left( 5-3|r|-\sqrt{-3 (1-|r|)^2+1} \right) &, \,0{.}5\le|r|\le 1{.}5 \\
			\frac{1}{3}\left( 1+\sqrt{-3|r|^2+1} \right) &, \,|r|< 0{.}5 \\
			0 &, \,\text{otherwise}
   \end{cases}
 \end{equation}
is employed, so that $\Phi_3$ has a width of $d_\Phi=3h$ as sketched in Fig.~\ref{fig:rod_discrete}. This ensures a good balance between numerical efficiency and smoothing properties~\cite{Kempe2012}. As an alternative the four-point version of Peskin~\cite{Peskin2002}
\begin{equation}
	 \label{eqn:dirac_peskin}
	 \Phi_4(r) = 
   \begin{cases}
			\frac{1}{8}\left( 5-2|r|-\sqrt{-7 + 12|r| - 4|r|^2} \right) &, \,1\le|r|\le 2 \\
			\frac{1}{8}\left( 3-2|r|+\sqrt{1 + 4|r| - 4|r|^2} \right) &, \,|r|< 1 \\
			0 &, \,\text{otherwise}
   \end{cases}
\end{equation}
will be considered as well in the validation below.\\ 
With the help of the regularized delta function $\delta_h$ a transfer of an arbitrary vector quantity $\vec{\varphi}$ from the Eulerian points $\vec{x}_{ijk}$ to the Lagrangian points $\vec{x}_l$ is accomplished by an interpolation via
\begin{equation}
	\label{eqn:interpolation}
	\vec{\varphi}(\vec{x}_l) = 
	\sum^{N_x}_{i=1} \sum^{N_y}_{j=1} \sum^{N_z}_{k=1}\vec{\varphi}(\vec{x}_{ijk}) \:\delta_h(\vec{x}_{ijk}-\vec{x}_l) \:h^3 \, ,
\end{equation}
e.g. to provide fluid velocities at the location of the interface $\Gamma$. The complementary operation is a transfer from Lagrangian to Eulerian points, often called spreading or regularization. It is defined by
\begin{equation}
	\label{eqn:regularization}
	\vec{\varphi}(\vec{x}_{ijk}) =
	\sum_{\;\vec{x}_{l}\,\in\,\Gamma_e} \vec{\varphi}(\vec{x}_l)\: \delta_h(\vec{x}_{ijk}-\vec{x}\vec{}_l) \:\Delta V_l
\end{equation}
and is commonly used to distribute the coupling force $\vec{f}_\IBM$ to the Eulerian grid used to solve the equations for the fluid. The width of the regularized delta function, $d_\Phi$, introduced in this section, and the thickness of the Lagrangian layer $d_\Lambda$ of the previous section are two different aspects of the discretization scheme. The width $d_\Lambda$ is required for the definition of appropriate forcing volumes $\Delta V_l$ associated to each forcing point. From a numerical point of view, this is the discrete realization of the support of the delta function $\delta_V$ in the continuous formulation of coupling force~\eqref{eqn:surface_to_volume_integrals}, and is uniquely defined by the discretization of the Eulerian grid. The second width $d_\Phi$ is an independent parameter and can be chosen ``arbitrarily'' by selecting a certain regularized delta function. It can be interpreted as the width of regularization regarding the spreading operation that serves to transfer momentum from the Lagrangian points to the Eulerian points. Due to the distributive nature of the momentum source $\vec{f}_\IBM$ regularization reduces or avoids jumps of $\vec{f}_\IBM$ on the Eulerian grid and, thus, prevents numerical oscillations. Interpolation an spreading are performed with the same delta function.

\subsection{Proposed coupling algorithm}
\label{sec:coupling scheme}
Temporal and spatial coupling of the Navier-Stokes equations~\eqref{eq:NSE} and the Cosserat rod equations~\eqref{eqn:Cosserat} are now combined to a partitioned solution approach. It is realized in a fully explicit manner, which is exempt from any global iteration between the fluid part and the structure part. The scheme presented in section~\ref{sec:resulting_scheme} is now detailed, with the following five steps executed once in each Runge-Kutta sub-step $r$. A compact overview of the proposed FSI scheme is provided in~\ref{sec:FSI algorithm}.

\begin{figure}[!tb]
 \centering
 \includegraphics[trim=0 0 0 0, clip, scale=1]{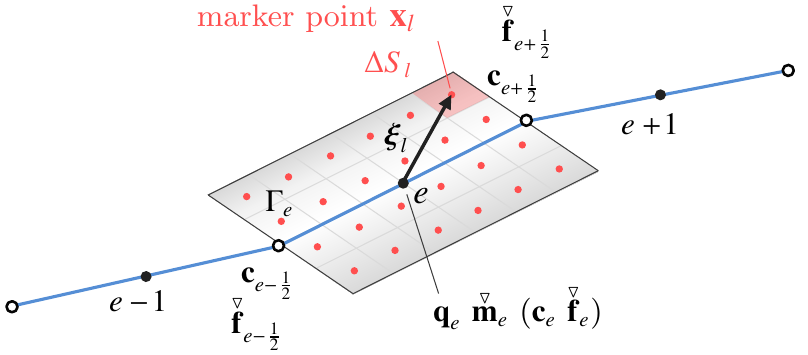}
 \caption{\label{fig:rod_fp}Discrete structural rod element $\Gamma_e\!\subset\!\Gamma$ represented by uniformly distributed Lagrangian marker points $\vec{x}_l$, each covering a surface area $\Delta S_l$. The vector $\vec{\xi}_l = \vec{x}_l - \vec{c}_e$ denotes the relative position of $\vec{x}_l$ with respect to the element center position $\vec{c}_e$. Due to the staggered spatial discretization the quaternions $\vec{q}_e$ and the external moments $\overset{\smalltriangledown}{\vec{m}}_e$ are given at the element center, while the center line positions $\vec{c}_{e\pm1/2}$ and the external forces $\overset{\smalltriangledown}{\vec{f}}_{e\pm1/2}$ are defined between two adjoining elements. }
\end{figure}

\subparagraph{1. Computation and interpolation of preliminary velocities.}
First, the preliminary velocity field $\tilde{\vec{u}}(\vec{x}_{ijk})$ is computed on the Eulerian grid points $\vec{x}_{ijk}$, according to Eqs.~\eqref{eqn:prelim_vel} and \eqref{eqn:RK_expl} using the Runge-Kutta scheme applied here. Thereafter, the values $\tilde{\vec{u}}(\vec{x}_l^{r-1})$ located at the Lagrangian marker points of the previous time level $\vec{x}_l^{r-1}$ (Fig.~\ref{fig:rod_fp}) are interpolated from the Eulerian grid points $\vec{x}_{ijk}$ according to Eq.~\eqref{eqn:interpolation}, so that
\begin{equation}
	\label{eqn:Dir_U_dis}
	\tilde{\vec{u}}(\vec{x}_l^{r-1}) = 
	\sum^{N_x}_{i=1} \sum^{N_y}_{j=1} \sum^{N_z}_{k=1}\tilde{\vec{u}}(\vec{x}_{ijk}) \:\delta_h(\vec{x}_{ijk}-\vec{x}_l^{r-1}) \:h^3 \, .
\end{equation}

\subparagraph{2. Communication of preliminary quantities to the structure solver.}
According to the direct forcing approach described in section~\ref{sec:direct forcing}, the values $\tilde{\vec{u}}$ on $\Gamma$ are used to determine the coupling force $\vec{f}_\IBM$ and thus are required to compute the fluid loads $\fext_\IBM$, $\mext_\IBM$ acting on the rod during motion. Therefore, the velocities $\tilde{\vec{u}}(\vec{x}_l^r)$ for $N_l$ marker points need to be transferred to the corresponding structure solver. In order to realize a coupling to Cosserat rods these can be replaced by integral quantities $\tilde{\vec{p}}_e$, $\tilde{\vec{l}}_e$ (see Eqs.~\eqref{eqn:fluid_force_quat2} and~\eqref{eqn:fluid_moment_quat2} below) for each rod element $e$. This drastically reduces the communication effort between the fluid and structure solver to 6 values per element, i.e. $6 N_\e$ per rod in total.

\subparagraph{3. Solving Cosserat rod equations modified by coupling terms.}
To realize the coupling to the surrounding fluid, the Cosserat rod equations~\eqref{eqn:Cosserat_quat_linear} are modified via the coupling terms~\eqref{eqn:coupling_terms_mod1} and \eqref{eqn:coupling_terms_mod2}. The corresponding discrete versions are given by
\begin{subequations}
\begin{align}
  \label{eqn:fext}
	\fext_{\IBM,e}\,\Delta Z &= - \sum_{\;\vec{x}_{l}\,\in\,\Gamma_e} \Delta m_l\;\vec{f}_{\IBM,\text{lin}}(\dot{\vec{c}},\dot{\vec{q}},\vec{q},t)  \\
	\mext_{\IBM,e}\,\Delta Z &= - \sum_{\;\vec{x}_{l}\,\in\,\Gamma_e} \Delta m_l\,\vec{\xi}_l \times \vec{f}_{\IBM,\text{lin}}(\dot{\vec{c}},\dot{\vec{q}},\vec{q},t)
\end{align}
\end{subequations}
for an individual rod element $\Gamma_\e \subset \Gamma$ represented by Lagrangian marker points $\vec{x}_l$. Each marker covers a Lagrangian fluid layer mass $\Delta m_l = \rho_\f \,h\,\Delta S_l$. Assuming a rigid body motion of an element, as discussed in section~\ref{sec:discretization_rod}, the fluid force~\eqref{eqn:fext} acting on $\Gamma_\e$ can be reformulated with $\vec{f}_{\IBM,\text{lin}}$~\eqref{eqn:lin_IBM_force} as
\begin{subequations}
\label{eqn:fluid_force}
\begin{gather}
  \label{eqn:fluid_force_quat}
	\fext_{\IBM,e}\,\Delta Z = -\left[ \frac{{\vec{p}_\Gamma} - \vec{p}^{r-1}_\Gamma}{t-t^{r-1}} + \frac{\vec{p}^{r-1}_\Gamma - \tilde{\vec{p}}}{2\alpha_r\,\Delta t} \right]_e \quad \text{with} \;\quad
	\vec{p}_{\Gamma,e} = \left[ \vec{q}\qm(m\,\dot{\vec{c}}\tra + \vec{\omega}\tra\times\vec{s}\tra)\qm\bar{\vec{q}} \,\right]_e \\
	\label{eqn:fluid_force_quat2}
	m_e = \sum_{\;\vec{x}_{l}\,\in\,\Gamma_e} \Delta m_l \, , \qquad 
	\vec{s}_{\text{\tiny{0}},e} = \sum_{\;\vec{x}_{l}\,\in\,\Gamma_e} \Delta m_l\,\vec{\xi}_{\text{\tiny{0}},l} \, , \qquad 
	\tilde{\vec{p}}_e = \sum_{\;\vec{x}_{l}\,\in\,\Gamma_e} \Delta m_l\,\tilde{\vec{u}}(\vec{x}_{l}^{r-1}) \, ,
\end{gather}
\end{subequations}
where the vector $\vec{p}$ designates the linear momentum of the fluid layer around an element $e$. Related quantities are the fluid layer mass~$m_e$, the static moment of the layer~$\vec{s}_{\text{\tiny{0}},e}$ and the preliminary linear momentum $\tilde{\vec{p}}_e$ as an integral measure of $\tilde{\vec{u}}(\vec{x}^r_l)$, computed in the previous step. These quantities can be precomputed before solving the rod equations. Due to the staggered spatial discretization of the rod, the external forces $\fext_\IBM$ are considered at nodes with half-index, i.e. $e+\frac{1}{2}=\frac{1}{2},\,...\,N_{e+\frac{1}{2}}$ (Fig.~\ref{fig:rod_fp}). Here, $\fext_{\IBM,e+\frac{1}{2}}$ is approximated by the mean value of the fluid forces of both adjoining elements, i.e. $\fext_{\IBM,e+\frac{1}{2}}= (\fext_{\IBM,e} + \fext_{\IBM,e+1})/2$. In a similar manner, the velocities $\dot{\vec{c}}_e$ are determined from the values at the element edges, i.e. $\dot{\vec{c}}_{e} = (\dot{\vec{c}}_{e+\frac{1}{2}} + \dot{\vec{c}}_{e-\frac{1}{2}})/2$. Analogous to the external fluid forces, the external moments acting on $\Gamma_e$ are obtained via
\begin{subequations}
\label{eqn:fluid_moment}
\begin{gather}
  \label{eqn:fluid_moment_quat}
	\mext_{\IBM,e}\,\Delta Z = -\left[ \frac{{\vec{l}_\Gamma} - \vec{l}^{r-1}_\Gamma}{t-t^{r-1}} + \frac{\vec{l}^{r-1}_\Gamma - \tilde{\vec{l}}}{2\alpha_r\,\Delta t} \right]_e \quad \text{with} \quad 
	\vec{l}_{\Gamma,e} = \left[ \vec{q}\qm( \vec{s}\tra\times\dot{\vec{c}}\tra + \mat{\mathcal{J}}\tra\cdot\vec{\omega}\tra )\qm\bar{\vec{q}} \,\right]_e \\
	\label{eqn:fluid_moment_quat2}
	\mat{\mathcal{J}}_{\text{\tiny{0}},e} = 0\oplus\!\!\! \sum_{\;\vec{x}_{l}\,\in\,\Gamma_e} \Delta m_l\,[\vec{\xi}_{\text{\tiny{0}},l}]^\top_\times \cdot [\vec{\xi}_{\text{\tiny{0}},l}]_\times \, , \qquad 
	\tilde{\vec{l}}_e = \sum_{\;\vec{x}_{l}\,\in\,\Gamma_e} \Delta m_l\,\vec{\xi}^n_{l}\times\tilde{\vec{u}}(\vec{x}^{r-1}_{l}) \, ,
\end{gather}
\end{subequations}
where $\vec{l}$ designates the angular momentum of the fluid layer. As for the linear momentum, the static moment of the layer~$\vec{s}_{\text{\tiny{0}},e}$, the quaternionic tensor of inertia $\mat{\mathcal{J}}_{\text{\tiny{0}},e}$ and the preliminary angular momentum $\tilde{\vec{l}}_e$ can be precomputed.\\
In step 3 of the direct forcing coupling scheme (Fig.~\ref{fig:temp_coupling}), the Cosserat rod equations~\eqref{eqn:Cosserat_quat} modified by $\fext_{\IBM,e}$ and $\mext_{\IBM,e}$ are solved implicitly for the new linear velocities of rod center line $\dot{\vec{c}}^{r}_{e-\frac{1}{2}}$ and the angular velocities $\vec{\omega}^{r}_e = 2\,\dot{\vec{q}}^r_e\!\qm\bar{\vec{q}}^r_e$ by means of the quaternions $\vec{q}^r_e$, $\dot{\vec{q}}^r_e$.
\subparagraph{4. Communication of coupling forces to the fluid solver.}
In the next step, the velocities $\dot{\vec{c}}^{r}_e=(\dot{\vec{c}}^r_{e+\frac{1}{2}} + \dot{\vec{c}}^r_{e-\frac{1}{2}})/2$ and $\vec{\omega}^{r}_e$ at the new Runge-Kutta time level are communicated to the fluid solver. Based on these velocities the corresponding interface velocity of a rod element $e$  is computed via 
\begin{equation}
	\label{eqn:u_gamma_discrete}
\textbf{u}^r_{\Gamma_e}(\vec{x}_l^{r-1}) = \dot{\vec{c}}^{r}_{e} + \vec{\omega}^{r}_e \times \vec{\xi}_l^{r-1} \,.
\end{equation}
With the preliminary velocities $\tilde{\vec{u}}(\vec{x}_l^{r-1})$ computed in step 1, the coupling force located at an individual Lagrangian point then is given by
\begin{equation}
	\label{eqn:direct_forcing_discrete}
	\bar{\vec{f}}_\IBM(\vec{x}_l^{r-1}) = \frac{\vec{u}^{r}_\Gamma(\vec{x}_l^{r-1}) - \tilde{\vec{u}}(\vec{x}_l^{r-1})}{2\alpha_r\,\Delta t} \, ,
\end{equation}
according to Eq.~\eqref{eqn:f_ibm_impl2}. Here, $\bar{\textbf{f}}_\IBM$ is formulated with the preliminary velocity $\tilde{\textbf{u}}\:(\textbf{x}^{r-1}_l)$ using the marker location $\textbf{x}^{r-1}_l$ at the old time level $r-1$, which amounts to a semi-implicit treatment of the coupling force.

\subparagraph{5. Spreading of coupling forces and reintegration of NSE.}
In a final step, the remaining equations of the fractional step scheme \eqref{eqn:helm+f}-\eqref{eqn:ibm_proj_1} are solved to obtain the new fluid velocity field $\textbf{u}^{r}$ and the pressure field $p^{r}$. Herein, the Helmholtz equation~\eqref{eqn:helm+f} includes $\vec{f}_\IBM(\vec{x}_{ijk})$, so that the fluid motion now is constraint by the immersed boundary $\Gamma$. Since the coupling forces $\bar{\vec{f}}_\IBM(\vec{x}_l^{r-1})$ computed in step 4 are only provided at the Lagrangian points $\vec{x}_l^{r-1}$, they are distributed to the Eulerian grid points $\vec{x}_{ijk}$ via the spreading operation~\eqref{eqn:regularization}, i.e.
\begin{equation}
	\label{eqn:Dir_f_dis}
	\bar{\vec{f}}_\IBM(\vec{x}_{ijk}) =
	\sum_{\;\vec{x}_{l}\,\in\,\Gamma_e} \bar{\vec{f}}_\IBM(\vec{x}_l^{r-1})\;\, \delta_h(\vec{x}_{ijk}-\vec{x}_l^{r-1}) \:\Delta V_l \,.
\end{equation}

\subsection{Numerical study of convergence}
\label{sec:Conv direct forcing}

\subsubsection{Test configuration}
\begin{figure}[!ht]
  \setlength{\unitlength}{1cm}
	
  {\begin{minipage}{0.60\txtw}

     \begin{tabular}{p{0.40\txtw} l}
       \hline
       \multicolumn{2}{l}{physical parameters:} \\
       $H = \SI{1}{m}$                               & channel height \\
       $\rho_\f = \SI{100}{kg/m^3}$                  & fluid density \\
       $\nu_\f = \SI{0{.}005}{m^2 / s}$              & kinematic viscosity \\
       $U = \SI{1}{m/s}$                             & shear velocity \\[0.3cm]
       \multicolumn{2}{l}{dimensionless quantities:} \\
       $Re_{\frac{H}{2}} = 100$                      &	Reynolds number \\
                                                     & $Re_{\frac{H}{2}} < Re_{\mathrm{crit}}=600$ \\[0.3cm]
       \multicolumn{2}{l}{numerical parameters:} \\
       $L_x \!=\! L_y \!=\! L_z = H$                 & domain size (cubic)	\\
       $N_x \!=\! N_y \!=\! N_z = 8\cdot 2^i$        & number of grid cells, $i\ge 0$ \\
       $\Delta x = 1/N_x$                            & step size of Eulerian grid \\
       $\Delta x_l = 2^{-i}/17$                      & step size of Lagrangian grid \\
       $\phi = \arctan(8/15)$                        & rotation of Lagrangian grid \\
       \hline
     \end{tabular}
		
  \end{minipage}}
 {\begin{minipage}{0.40\txtw}
   \centering
	 \includegraphics[trim=0 0 0 0, clip, scale=1]{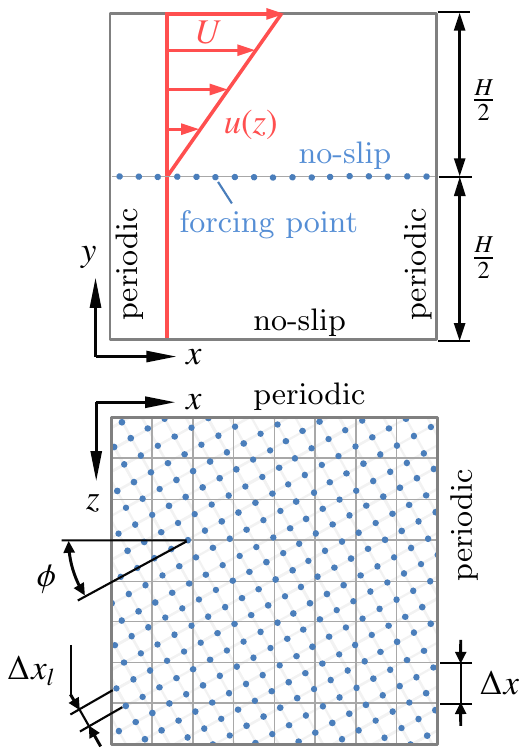}
 \end{minipage}}
  
 \caption{\label{fig:Couette}Planar shear flow driven by a constant shear velocity $U$ at the top of the fluid domain. The no-slip condition is imposed by a layer of uniform distributed forcing points located at $y=H/2$. Below this plane the fluid remains at rest. The present setup is used to analyze the convergence behavior of the direct forcing approach employed at the layer of forcing points. }
\end{figure}
The convergence behavior of the proposed direct forcing IBM is assessed by a simple steady planar shear flow. The physical parameters of the problem are listed in Fig.~\ref{fig:Couette}. The computational domain extends over a height $H$, with a no-slip condition $\vec{u}=(0,0,0)^\top$ at the bottom and a moving wall with $\vec{u}=(U,0,0)^\top$ at the top. Positioning an interface $\Gamma$ at $y = H/2$ mimics a solid structure of vanishing thickness and is addressed as immersed wall here. It results in the exact solution for the $x$-component of the velocity 
\begin{equation}
  u(y) = 
  \begin{cases} 
   \; 0 \,, & y \le \frac{H}{2} \\
   \; U\,(2y-H)\,, & \frac{H}{2} < y \le H \, ,
  \end{cases}
\end{equation}
while the pressure is uniform, $p=\text{const.}$ The boundary conditions at $y=0$ and $y=H$ are imposed on the Eulerian grid as usual. The no-slip condition at $y=H/2$, instead, is imposed by a layer of forcing points arranged as a two-dimensional Cartesian grid of spacing $\Delta x_l$. The layer is rotated by an angle of $\phi=\arctan(8/15)$ around the $y$-axis, to achieve a high degree of variation between the arrangement of the forcing points and the discretization of the fluid domain, as it is the case for freely movable structures.\\
Due to the kink in the velocity profile at $y=H/2$ the hydrodynamic stresses $\mat{\sigma}$ are different on both sides of the interface $\Gamma$. According to Eq.~\eqref{eqn:jump_surface_force}, the jump in $\mat{\sigma}$ is associated to a fluid load acting on $\Gamma$. For the present configuration this load simplifies to 
\begin{equation}
   \int_{\Gamma} \mat{\sigma}^+\cdot\vec{n} \:\text{d} S - {\int_{\Gamma}  \mat{\sigma}^-\cdot\vec{n} \d} S \, = \, \tau_\mathrm{w} H^2\, \vec{e}_x
\end{equation}
and, thus, is solely determined by the shear stress $\tau_w = \;\, 2\, \rho_\f \nu_\f \, U/H$ in the streamwise direction acting on the upper side of the immersed wall. In terms of the present direct forcing IBM, each forcing point $\vec{x}_l$ acts against this shear to impose the no-slip condition at $y=H/2$. Using Eq.~\eqref{eqn:fext}, with $\vec{u}_{\Gamma}=\vec{0}$ at the immersed wall, the shear force acting on $\Gamma$ can be approximated numerically by
\begin{equation}
  \label{eqn:wall shear stress}
	\tau_\mathrm{w} H^2\vec{e}_x \;\approx\;\, \vec{f}_{\mathrm{w}} = \sum^3_{r=1}\!\sum_{\;\vec{x}_{l}\,\in\,\Gamma} \Delta m_l\,\tilde{\vec{u}}(\vec{x}_{l}) / \Delta{t} \, ,
\end{equation}
with $\Delta m_l = \rho_\f \, \Delta x\,(\Delta x_l/H)^2$ for the present spatial arrangement of the forcing points. Since the components of the fluid loads in $y$- and $z$-direction vanish, the relation between the wall shear and the direct forcing is given by $\tau_\mathrm{w}\!\approx\vec{f}_{\mathrm{w}}\cdot\vec{e}_x /H^2$. The following convergence analysis bases on the relative error between the theoretical value of the wall shear and the numerical approximation
\begin{equation}
 \label{eqn:epsilon}
 \epsilon = \vec{f}_{\mathrm{w}}\cdot\vec{e}_x / (\tau_\mathrm{w} H^2) - 1 \, .
\end{equation}

\subsubsection{Numerical parameters}
To determine numerically the spatial and the temporal convergence rate, the error was computed over a wide range of grid step sizes $\Delta{x} \in \{1/8,1/16,1/32,1/64\}$ and Courant numbers $\CFL \in \{1,0{.}5,0{.}25,...,1/2^{6},0{.}01\}$ with $\CFL=U \Delta{t}/ \Delta{x}$. The spatial distribution of the forcing points, controlled by $\Delta x_l = \mathrm{const.}$, was adapted to the step size of the Eulerian grid, so that $\Delta x_l/\Delta x=8/17\approx 0{.}5$ for any $\Delta x$. Besides the temporal and spatial resolution the error is also influenced by the delta function employed for interpolation of the preliminary velocity $\tilde{\vec{u}}$ to the Lagrangian positions $\vec{x}_l$, as well for spreading the IBM force $\vec{f}_{\IBM}$ to the Eulerian grid points $\vec{x}_{ijk}$. Two different delta functions are tested in the present study, which are frequently employed in continuous direct forcing schemes~\cite{Peskin2002,Uhlmann2005,Yang2009,Kempe2012,Tschisgale2016}. One is the three-point delta function $\Phi_3$ of \mbox{Roma \textit{et al.}~\cite{Roma1999}} defined in Eq.~\eqref{eqn:dirac_roma}. The other is the four-point delta function $\Phi_4$ proposed by Peskin~\cite{Peskin2002} given in Eq.~\eqref{eqn:dirac_peskin}. 

\subsubsection{Results and discussion}
\begin{figure}[!tb]
 \centering
 \includegraphics[trim=0 0 0 0, clip, scale=1]{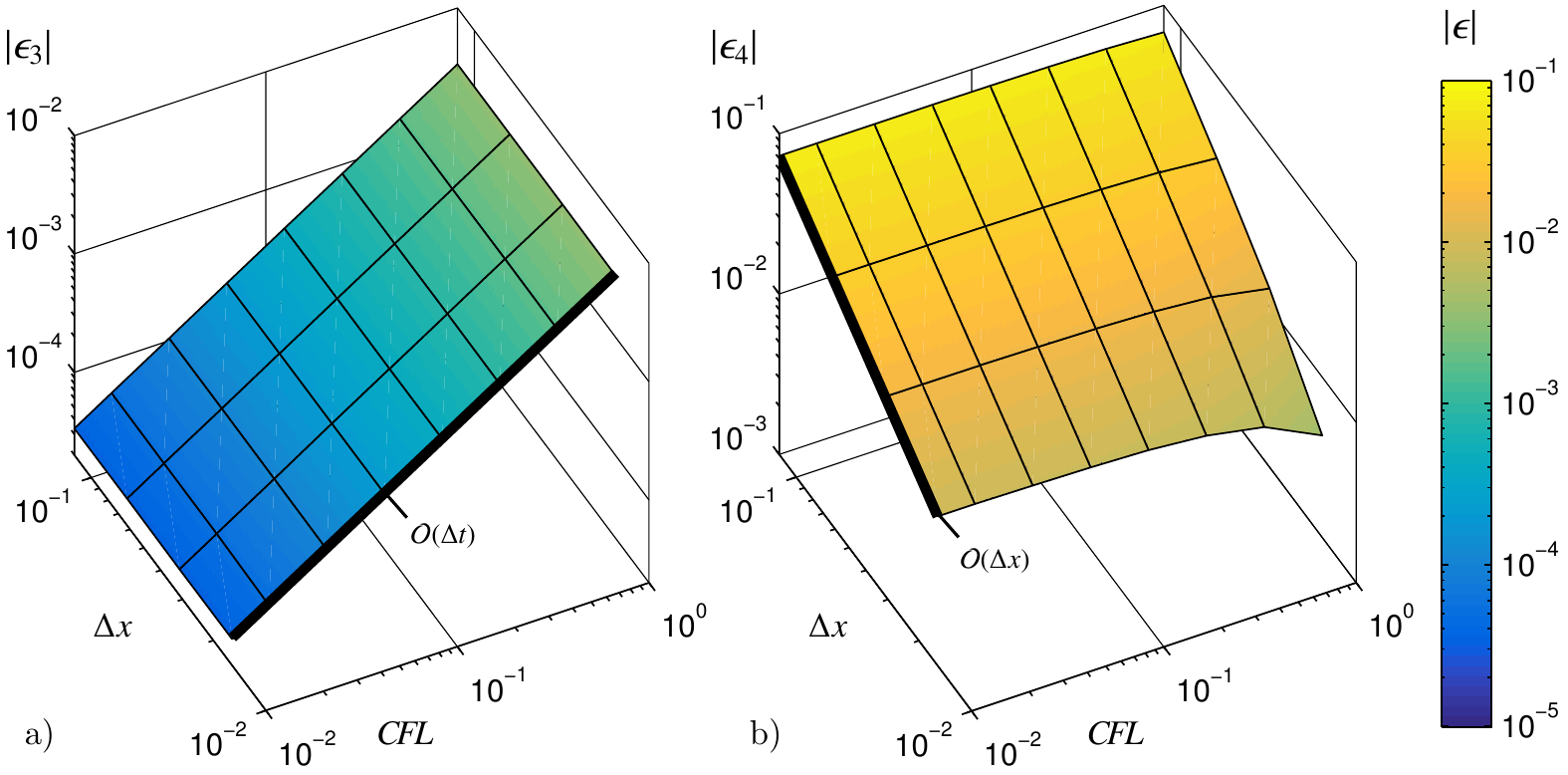}
 \caption{\label{fig:Couette_convergence}Temporal and spatial convergence of the present direct forcing--IBM in a range of $\Delta x \times \CFL = \{1/8,1/16,1/32,1/64\}\times\{1,0{.}5,0{.}25,...,1/2^6,0{.}01\}$. Shown are the relative errors $\epsilon_3$ and $\epsilon_4$ according to Eq.~\eqref{eqn:epsilon} obtained from simulations with the 3-point delta function $\Phi_3$ and the 4-point delta function $\Phi_4$, respectively.}
\end{figure}

The convergence behavior obtained for $\Phi_3$ and $\Phi_4$ in the given range of spatial and temporal resolutions is shown in Fig.~\ref{fig:Couette_convergence} using $\epsilon$ from Eq.~\eqref{eqn:epsilon}. Selected values of the corresponding errors $\epsilon_3$ and $\epsilon_4$, respectively, are provided in Table~\ref{tab:Couette}.
\begin{table}[t]

 \setlength{\tabcolsep}{3pt}
 \centering
 \begin{tabular}{P{0.09\txtw} | P{0.09\txtw} | P{0.09\txtw} | P{0.17\txtw} | P{0.17\txtw}}
 $1/\Delta{x}$ & $1/\CFL$ & $1/\Delta{t}$ & $|\epsilon_3|$ & $|\epsilon_4|$ \\ \hlineB{2}
  8			& 64			& 512				& \SI{5{.}248e-5}{} & \SI{7{.}275e-2}{}	\\
  16		& 64			& 1024			& \SI{5{.}237e-5}{} & \SI{3{.}509e-2}{}	\\
  32		& 64			& 2048			& \SI{5{.}196e-5}{} & \SI{1{.}731e-2}{}	\\
  64		& 64			& 4096			& \SI{5{.}040e-5}{} & \SI{8{.}510e-3}{}	\\
  64		& 32			& 2048			& \SI{1{.}022e-4}{} & \SI{8{.}433e-3}{}	\\
  64		& 16			& 1024			& \SI{2{.}086e-4}{} & \SI{8{.}295e-3}{}	\\
  64		& 8				& 512				& \SI{4{.}194e-4}{} & \SI{8{.}026e-3}{}	\\
 \end{tabular}

\caption{\label{tab:Couette}Relative errors $\epsilon$ of the wall shear stress $\tau_\mathrm{w}$ for selected grid step sizes $\Delta x$ and time step sizes $\Delta t$ ($\CFL=U\Delta t/\Delta x$). Listed are the errors $\epsilon_3$ and $\epsilon_4$ obtained with the 3-point delta function $\Phi_3$ and the 4-point delta function $\Phi_4$, respectively. }
\end{table}
Obviously, the convergence behavior is quite different for the two delta functions selected. While the three-point version exhibits a first order convergence in time for any spatial discretization, the four-point version shows the opposite behavior, i.e. a first order convergence in space for any $\CFL$ number. This is explained as follows:
The total numerical error of the present direct forcing IBM can be traced back to two sources. On one hand, the temporal part of the error results from the direct forcing approach which is used to estimate the amplitude of the coupling force at each forcing point. Due to its time splitting the method exhibits first order accuracy in time $\mathcal{O}(\Delta{t})$. On the other hand, the spatial part of the total error stems from the delta functions used for interpolation and spreading. While an evaluation of numerical errors for the spreading operation at least seems to be ambitious, the accuracy of an interpolation by means of delta functions is well understood. By increasing the width of support additional constraints can be incorporated, so that higher moments of $\Phi$ and its smoothness are conserved~\cite{Peskin2002,Yang2009}. As stated in \cite{Liu2012}, the moment order controls the accuracy in the low frequency range, while the smoothing order suppresses a possible Gibbs phenomenon that may corrupt convergence. A simple two-point linear hat-function, for example, exhibits a discontinuity in its first derivative that often leads to spatial oscillations in the solution~\cite{Kajishima2016}.\\
In fact, the convergence rate of the approximation depends on both, the smoothness of the approximating function as well as the smoothness of the function to be approximated. In the present IBM framework $\mat{\sigma}$ exhibits a jump at the interface $\Gamma$, so that $\vec{u}(\vec{x})$ is not differentiable at $\vec{x}\in\Gamma$.
As a consequence, the rate of spatial convergence reduces to $\mathcal{O}(\Delta{x})$ for any delta function, regardless of the support of $\Phi$~\cite{Beyer1992,Peskin2002}. Moreover it turns out, that a wider support increases the spatial error compared to a more narrow delta function. As shown in Fig.~\ref{fig:Couette_convergence}, the errors obtained for $\Phi_4$ with four-point support are increased by at least one order of magnitude compared to $\Phi_3$ with three-point support. The total error $\epsilon_4$ is dominated by spatial interpolation and spreading errors of order $\mathcal{O}(\Delta{x})$, while temporal errors, resulting from the direct forcing approach, are much smaller and just not recognizable. On the contrary, for $\Phi_3$ spatial errors are negligibly small compared to the temporal splitting error, so that $\epsilon_3$ mainly converges with $\mathcal{O}(\Delta{t})$.
In the present work, the three-point function of Roma \textit{et al.}~\cite{Roma1999} is preferred as it constitutes a good balance between accuracy, numerical efficiency and smoothing properties. 

\subsection{Validation and results}
\label{sec:CFSI_results}

\subsubsection{FSI problem of Wall and Ramm}

\begin{figure}[!ht]
  \setlength{\unitlength}{1cm}
	
  {\begin{minipage}{0.45\txtw}

    \begin{tabular}{p{0.55\txtw} l}
     \hline
     \multicolumn{2}{l}{fluid properties:} \\
     $L_x = \SI{19{.}5}{\centi m}$                    & channel length \\
     $L_y = \SI{13}{\centi m}$                        & channel height \\
     $\rho_\f = \SI{1{.}18e-3}{g/{\centi m}^3}$       & fluid density \\
     $\eta_\f = \SI{1{.}82e-4}{g/(\centi m s)}$       & dyn. viscosity \\
     $U = \SI{51{.}3}{\centi m/s}$                    & bulk velocity \\[0.3cm]
     \multicolumn{2}{l}{structure properties:} \\
     $W = \SI{1}{\centi m}$                           & square width \\
     $L = \SI{4}{\centi m}$                           & rod length \\
     $T = \SI{0{.}06}{\centi m}$                      & rod thickness \\
     $\rho_\mathrm{s} = \SI{0{.}1}{g/{\centi m}^3}$   & structure density \\
     $\nu_\mathrm{s} = \SI{0{.}35}{}$                 & Poisson ratio \\
     $E_\s = \SI{2{.}5e6}{g/(\centi m s^2)}$          & Young modulus \\
     $k_\mathrm{s} = 5/6$                             & shear correction \\[0.3cm]
     \multicolumn{2}{l}{dimensionless quantities:} \\
     $Re_W \approx 333$                               &	Reynolds number \\
     $\rho_\mathrm{s}/\rho_\f \approx 84.7$           & density ratio \\
     \hline
    \end{tabular}

 \end{minipage}}
 {\begin{minipage}{0.55\txtw}
  \centering
  \includegraphics[trim=0 0 0 0, clip, scale=1]{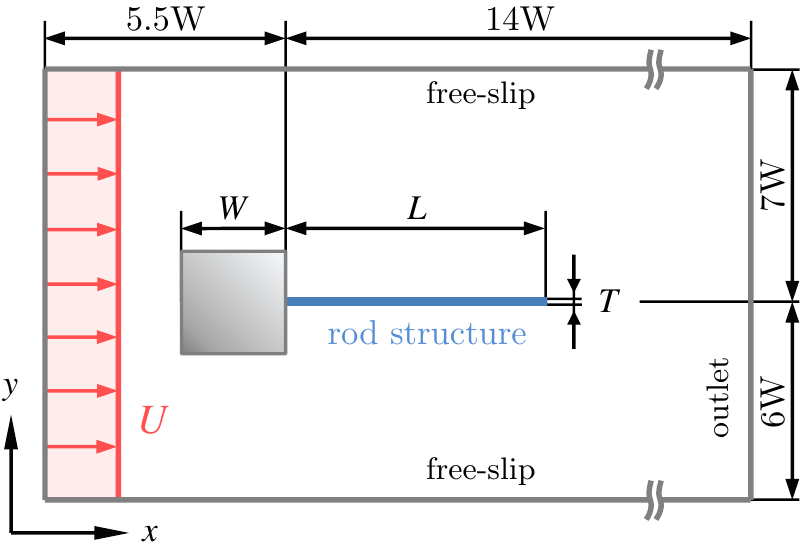}
 \end{minipage}}
  
 \caption{\label{fig:Dettmer}Setup according to Wall and Ramm~\cite{Wall1998, Wall1999} (not to scale). A slender elastic rod is mounted on an immobile square shaped obstacle submerged into a uniform flow. At the present Reynolds number of $Re_W\!=\!UW/\nu_\f \approx 333$ regular vortices are shed from the square and excite a periodic oscillation of the rod. }
\end{figure}
\begin{figure}[t]
 \setlength{\unitlength}{1cm}

 {\begin{minipage}{0.50\txtw}

  \setlength{\tabcolsep}{3pt}
  \begin{tabular}{c | c | c | c | c | c}
   \bf level & $\Delta{t}/10^{-4}\SI{}{s}$ & $\Delta{x}/10^{-1}\SI{}{cm}$ & $L/\Delta{x}$ & $N$ & $N_\mathrm{e}$ \\ \hlineB{2}
   $4h$ & 1				& 0{.}25		&  160	&  405600 	&	 20 \\ \cdashline{1-6}[2pt/2pt]
   $2h$ & 0{.}5		& 0{.}125		&  320	&  1622400 	&	 40 \\ \cdashline{1-6}[2pt/2pt]
   $h$  & 0{.}25	& 0{.}0625	&  640	&  6489600 	&	 80 \\
  \end{tabular}
	
 \end{minipage}}
{\begin{minipage}{0.50\txtw}
 \centering
 \includegraphics[trim=0 0 0 0, clip, scale=1]{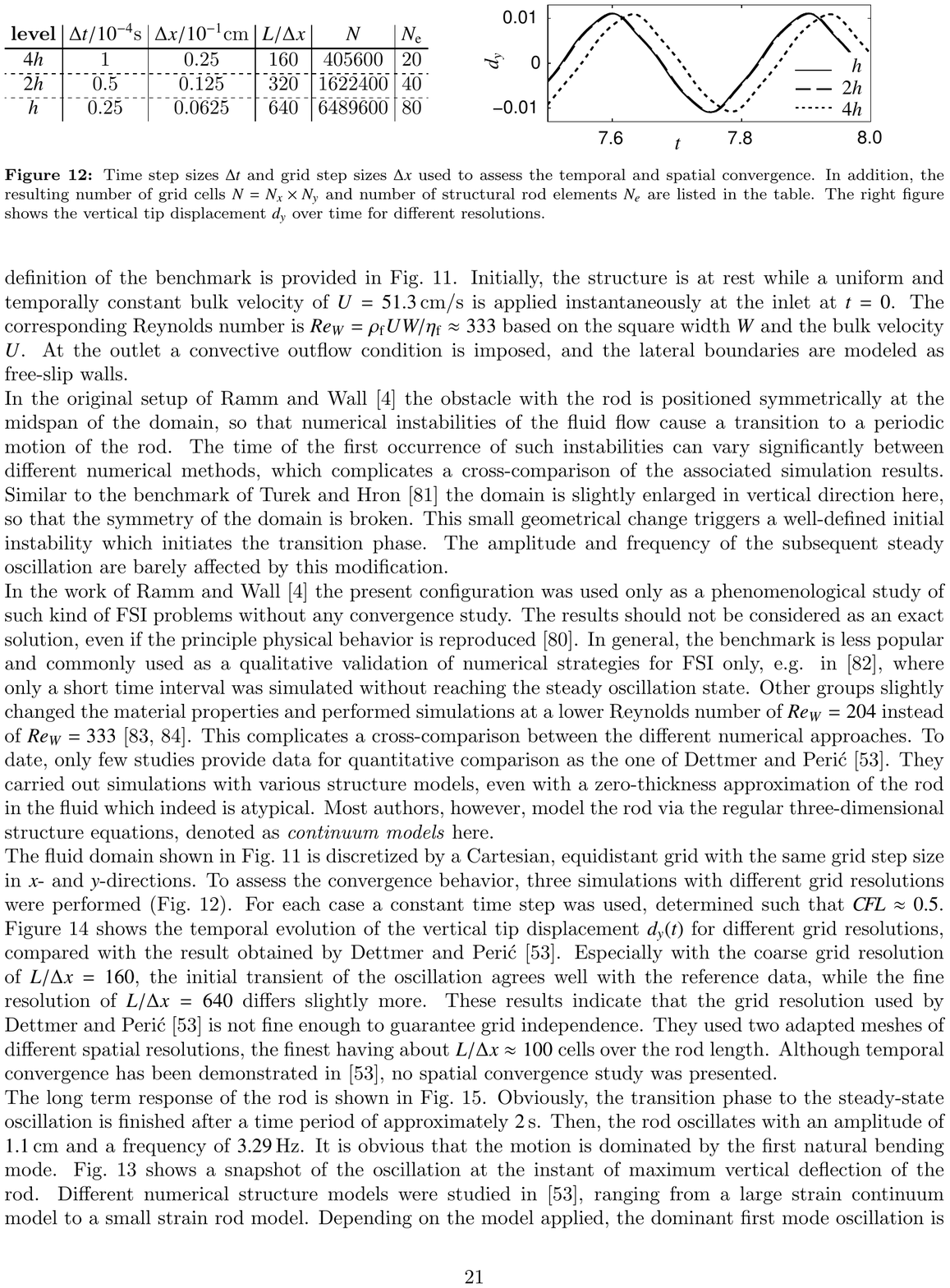}
\end{minipage}}

\caption{\label{fig:Dettmer_2}Time step sizes $\Delta{t}$ and grid step sizes $\Delta{x}$ used to assess the temporal and spatial convergence. In addition, the resulting number of grid cells $N = N_x \times N_y$ and number of structural rod elements $N_e$ are listed in the table. The right figure shows the vertical tip displacement~$d_y$ over time for different resolutions. }
\end{figure}
\begin{figure}[!tb]
  \centering
  \includegraphics[trim=0 0 0 0, clip, scale=1]{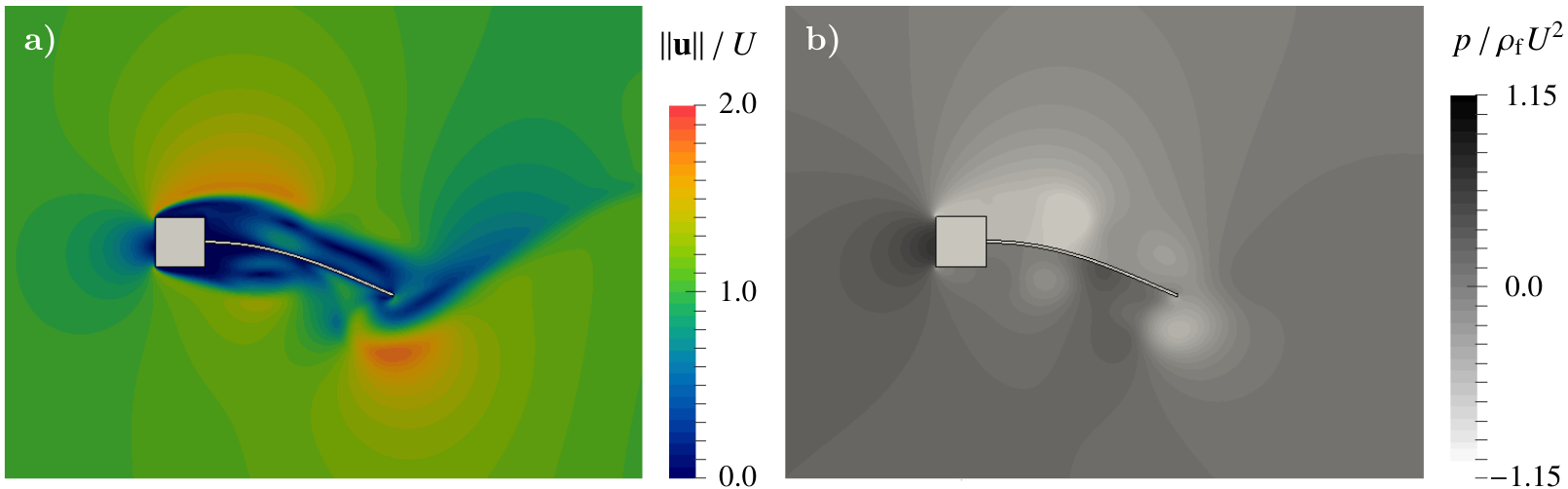} 
  \caption{\label{fig:Dettmer_5_contour}Instantaneous solution of the FSI problem according to Ramm and Wall at the time of maximum vertical deflection of the rod. a) Contour plot of instantaneous normalized velocity magnitude $\|\vec{u}\|/U$, b) normalized pressure field $p/\rho_\f U^2$. Clearly visible are the suction side (negative pressure, light gray) and the pressure side (positive pressure, dark gray), which generate a positive lift force on the structure. The light pressure region at the tip indicates the growth of an individual vortex.}
\end{figure}
\begin{figure}[!tb]
  \centering
  \includegraphics[trim=0 0 0 0, clip, scale=1]{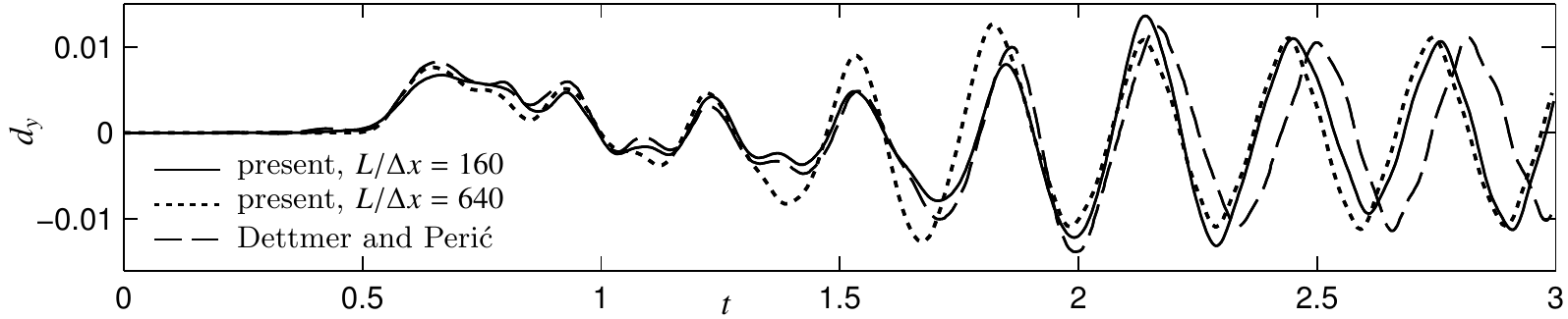}
  \caption{\label{fig:Dettmer_3_results}Vertical tip displacement $d_y$ during the initial transition phase. The results obtained for the coarse resolution ($L/\Delta{x} = 160$) and the fine resolution ($L/\Delta{x} = 640$) are compared to the data of Dettmer and Peri\'{c}~\cite{Dettmer2006} (picked case: fine grid, small strain continuum structure model). }
\end{figure}
\begin{figure}[!tb]
  \centering
  \includegraphics[trim=0 0 0 0, clip, scale=1]{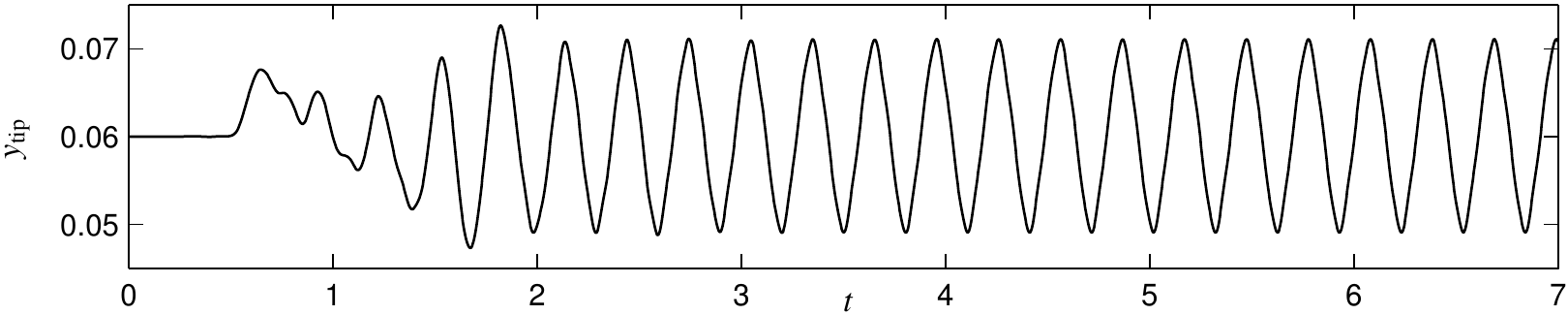}
  \caption{\label{fig:Dettmer_4_results}Long-term response of the rod shown for the vertical tip position $y_\mathrm{tip}$. After an initial transition phase of approximately \SI{2}{s} the rod oscillates with a constant amplitude of~\SI{1{.}1}{\centi m} and a frequency of~\SI{3{.}29}{Hz}. }
\end{figure}

The commonly used FSI benchmark of Turek and Hron~\cite{Turek2006} is an improved version of the FSI problem proposed by Ramm and Wall~\cite{Wall1998, Wall1999} several years earlier. Both configurations base on the same physical phenomenon, a vortex-induced oscillation of a flexible rod in the wake of an immobile obstacle in laminar flow. Besides different material parameters for the fluid and the structure, an alternative shape of the obstacle is used, a square instead of a circle. Moreover, the thickness of the rod is significantly smaller, which is better suited to validate the present IBM using a zero-thickness representation of the rod. The definition of the benchmark is provided in Fig.~\ref{fig:Dettmer}. Initially, the structure is at rest while a uniform and temporally constant bulk velocity of $U = \SI{51{.}3}{\centi m/s}$ is applied instantaneously at the inlet at $t=0$. The corresponding Reynolds number is $Re_W = \rho_\f UW/\eta_\f \approx 333$ based on the square width $W$ and the bulk velocity $U$. At the outlet a convective outflow condition is imposed, and the lateral boundaries are modeled as free-slip walls.\\
In the original setup of Ramm and Wall~\cite{Wall1998} the obstacle with the rod is positioned symmetrically at the midspan of the domain, so that numerical instabilities of the fluid flow cause a transition to a periodic motion of the rod. The time of the first occurrence of such instabilities can vary significantly between different numerical methods, which complicates a cross-comparison of the associated simulation results. Similar to the benchmark of Turek and Hron~\cite{Turek2006} the domain is slightly enlarged in vertical direction here, so that the symmetry of the domain is broken. This small geometrical change triggers a well-defined initial instability which initiates the transition phase. The amplitude and frequency of the subsequent steady oscillation are barely affected by this modification.\\
In the work of Ramm and Wall~\cite{Wall1998} the present configuration was used only as a phenomenological study of such kind of FSI problems without any convergence study. The results should not be considered as an exact solution, even if the principle physical behavior is reproduced~\cite{Wall1999}. In general, the benchmark is less popular and commonly used as a qualitative validation of numerical strategies for FSI only, e.g. in~\cite{Steindorf2002}, where only a short time interval was simulated without reaching the steady oscillation state. Other groups slightly changed the material properties and performed simulations at a lower Reynolds number of $Re_W=204$ instead of $Re_W=333$~\cite{Xia2008, Huebner2004}. This complicates a cross-comparison between the different numerical approaches. To date, only few studies provide data for quantitative comparison as the one of Dettmer and Peri\'{c}~\cite{Dettmer2006}. They carried out simulations with various structure models, even with a zero-thickness approximation of the rod in the fluid. Most authors, however, model the rod via the regular three-dimensional structure equations, denoted as~\textit{continuum models} here.\\
The fluid domain shown in Fig.~\ref{fig:Dettmer} is discretized by a Cartesian, equidistant grid with the same grid step size in $x$- and $y$-directions. To assess the convergence behavior, three simulations with different grid resolutions were performed (Fig.~\ref{fig:Dettmer_2}).
For each case a constant time step was used, determined such that $\CFL \approx 0{.}5$. Figure~\ref{fig:Dettmer_3_results} shows the temporal evolution of the vertical tip displacement $d_y(t)$ for different grid resolutions, compared with the result obtained by Dettmer and Peri\'{c}~\cite{Dettmer2006}. Especially with the coarse grid resolution of $L/\Delta{x} = 160$, the initial transient of the oscillation agrees well with the reference data, while the fine resolution of $L/\Delta{x} = 640$ differs slightly more. These results indicate that the grid resolution used by Dettmer and Peri\'{c}~\cite{Dettmer2006} is not fine enough to guarantee grid independence. They used two adapted meshes of different spatial resolutions, the finest having about $L/\Delta{x} \approx 100$ cells over the rod length. Although temporal convergence has been demonstrated in~\cite{Dettmer2006}, no spatial convergence study was presented.\\
The long term response of the rod is shown in Fig.~\ref{fig:Dettmer_4_results}. Obviously, the transition phase to the steady-state oscillation is finished after a time period of approximately \SI{2}{s}. Then, the rod oscillates with an amplitude of \SI{1{.}1}{cm} and a frequency of \SI{3{.}29}{Hz}. It is obvious that the motion is dominated by the first natural bending mode. Fig.~\ref{fig:Dettmer_5_contour} shows a snapshot of the oscillation at the instant of maximum vertical deflection of the rod. Different numerical structure models were studied in~\cite{Dettmer2006}, ranging from a large strain continuum model to a small strain rod model. Depending on the model applied, the dominant first mode oscillation is superposed by a second mode of higher frequency. The displacement plot in Fig.~\ref{fig:Dettmer_3_results} shows that second mode oscillations also occur with the present Cosserat rod model ($t>\SI{2{.}5}{s}$), but with smaller amplitudes compared to the small strain continuum model of~\cite{Dettmer2006}. Tab.~\ref{tab:Dettmer_1} provides a cross-comparison between the present IBM and selected numerical approaches from the literature.
Each numerical approach reproduces the expected periodic behavior of the coupled system. However, deviations in the order of 10\% can be observed for the oscillation amplitude and frequency. H\"{u}bner \textit{et al.}~\cite{Huebner2004} showed how different stable periodic solutions can arise when different initial conditions are imposed. In the present study it was observed that the spatial resolution has a noticeable impact on the dynamic behavior. A very fine spatial discretization must be used to reach convergence, especially with regard to the initial transient phase.
\begin{table}[H]
						
 \setlength{\tabcolsep}{3pt}
 \small
 \centering
 
 \begin{tabular}{p{0.13\txtwt} | P{0.15\txtwt} | P{0.15\txtwt} | P{0.09\txtwt} : P{0.09\txtwt} | P{0.16\txtwt} | P{0.08\txtwt} : P{0.15\txtwt}}
  \bf numerical & IBM & IBM & IBM & ALE & ALE & \multicolumn{2}{c}{ALE} \\
  \bf method 		& \small present & 
  \small Gilmanov & 
  \multicolumn{2}{c|}{\small Kamensky } & 
  \small Baudille \& & 
  \multicolumn{2}{c}{\small Dettmer \& }
  \\[-0.1cm]  &&
  \textit{et al.}~\cite{Gilmanov2015} & 
  \multicolumn{2}{c|}{\textit{et al.}~\cite{Kamensky2015}} & 
  Biancolini~\cite{Baudille2008}& 
  \multicolumn{2}{c}{Peri\'{c}~\cite{Dettmer2006}}
  \\ \cdashline{1-8}[1pt/1pt]
  \specialcelll{\bf structure \\ \bf model} & 
  \specialcellc{Cosserat rod \\ \footnotesize (zero-thickness)}	&
  shell & shell & shell & continuum & beam & 
  \specialcellc{continuum \\ \footnotesize (small strain)} \\ \hlineB{2.5}
  {\bf ampl.} (\SI{}{cm}) & \bf 1{.}10 & 1{.}00 - 1{.}10 & 1{.}3 & $\approx 1$ 	& $\approx 1$	&	1{.}24	& 1{.}29 \\ \cdashline{1-8}[1pt/1pt]
  {\bf freq.} (\SI{}{Hz})				& \bf 3{.}29 & 3{.}2 & 3{.}2 & 3{.}2 				& 3{.}18			& 3{.}08	& 2{.}96
\end{tabular}

\caption{\label{tab:Dettmer_1}Oscillation amplitudes and frequencies of vertical tip-displacement $d_y$ obtained by different authors using either an IBM with a fixed background grid or an ALE method with a moving adapted mesh. Moreover, different structure models were applied in these works, ranging from non-reduced three-dimensional continuum models to one-dimensional rod models. Each structure model is formulated geometrically exact and thus is able to represent large rod deflections. }
\end{table}

\subsubsection{Flexible rod in cross flow}
\label{sec:Luhar}

\begin{figure}[!ht]
  \setlength{\unitlength}{1cm}
	
  {\begin{minipage}{0.4\txtw}

    \begin{tabular}{p{0.52\txtw} l}
		  \hline
      $g = \SI{981}{\centi m/s^2}$                       & grav. acceleration\\[0.3cm]
      \multicolumn{2}{l}{fluid properties (water):} \\
      $L_x = \SI{20}{\centi m}$                          & channel length \\
      $L_y = \SI{16}{\centi m}$                          & channel height \\
      $L_z = \SI{16}{\centi m}$                          & channel width \\
      $\rho_\f = \SI{1}{g/{\centi m}^3}$                 & fluid density \\
      $\nu_\f = \SI{0{.}01}{\centi m^2/s}$               & kin. viscosity \\
      $U = 3{.}6 \, \ldots \, 32 \, \SI{}{\centi m/s}$   & bulk velocity \\[0.3cm]
      \multicolumn{2}{l}{structure properties (foam material):} \\
      $L = \SI{5}{\centi m}$                             & rod length \\
      $W = \SI{1}{\centi m}$                             & rod width \\
      $T = \SI{0{.}2}{\centi m}$                         & rod thickness \\
      $\rho_\mathrm{s} = \SI{0{.}67}{g/{\centi m}^3}$    & structure density \\
      $\nu_\mathrm{s} = \SI{0{.}4}{}$                    & Poisson ratio \\
      $E_\s = \SI{50}{N/\centi m^2}$                     & Young modulus \\[0.3cm]
      \multicolumn{2}{l}{dimensionless quantities:} \\
      $Re_L = 360 \, \ldots \, 3200$                     &	Reynolds number \\
      $\rho_\mathrm{s}/\rho_\f = 0{.}67$                 & density ratio \\
			\hline
    \end{tabular}
	
  \end{minipage}}
  {\begin{minipage}{0.6\txtw}
   \centering
   \includegraphics[trim=0 0 0 0, clip, scale=1]{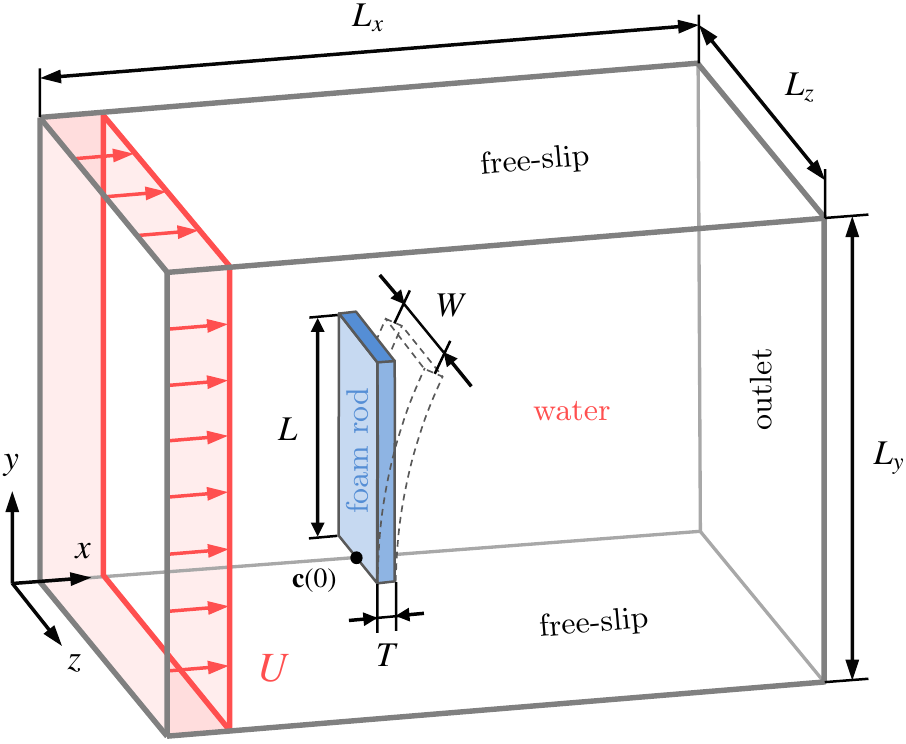}
  \end{minipage}}
  
 \caption{\label{fig:Luhar}Benchmark configuration of a flexible rod made subjected to a uniform cross flow according to the experimental work of Luhar and Nepf~\cite{Luhar2011} (drawing not to scale).}

\end{figure}
\begin{figure}[!tb]
  \centering
  \includegraphics[trim=0 0 0 0, clip, scale=1]{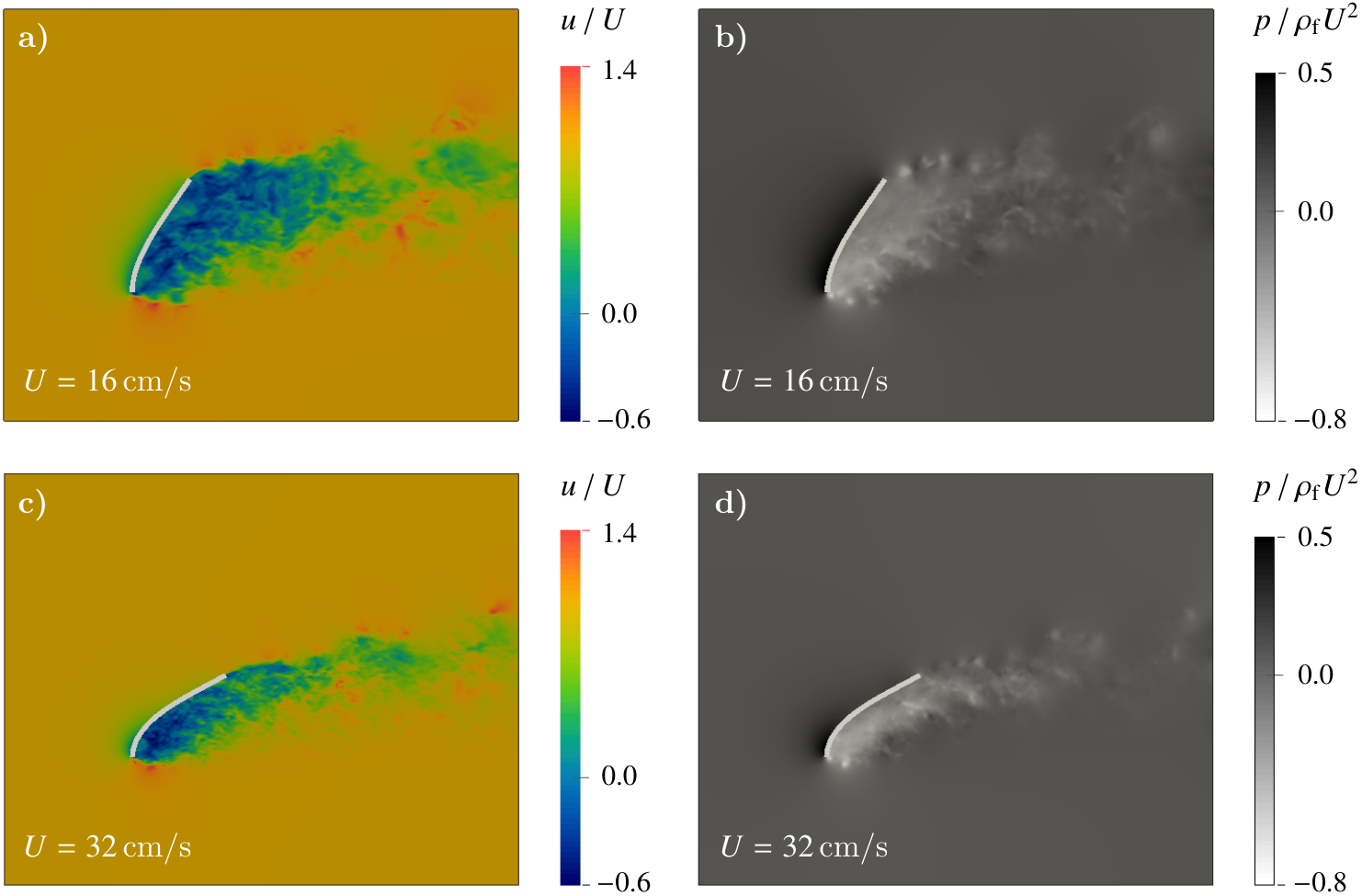}
  \caption{\label{fig:Luhar_1_contour}Instantaneous flow around the rod perpendicular to the mean flow in the center plane, $z=\SI{8}{\centi m}$, for two bulk velocities $U$. a) Streamwise velocity component $u$ for $U = \SI{16}{\centi m/s}$. b) Pressure field for $U = \SI{16}{\centi m/s}$. c) Streamwise velocity component $u$ for $U = \SI{32}{\centi m/s}$. d) Pressure field for $U = \SI{32}{\centi m/s}$.}
\end{figure}
\begin{figure}[H]
  \centering
  \includegraphics[trim=0 0 0 0, clip, scale=1]{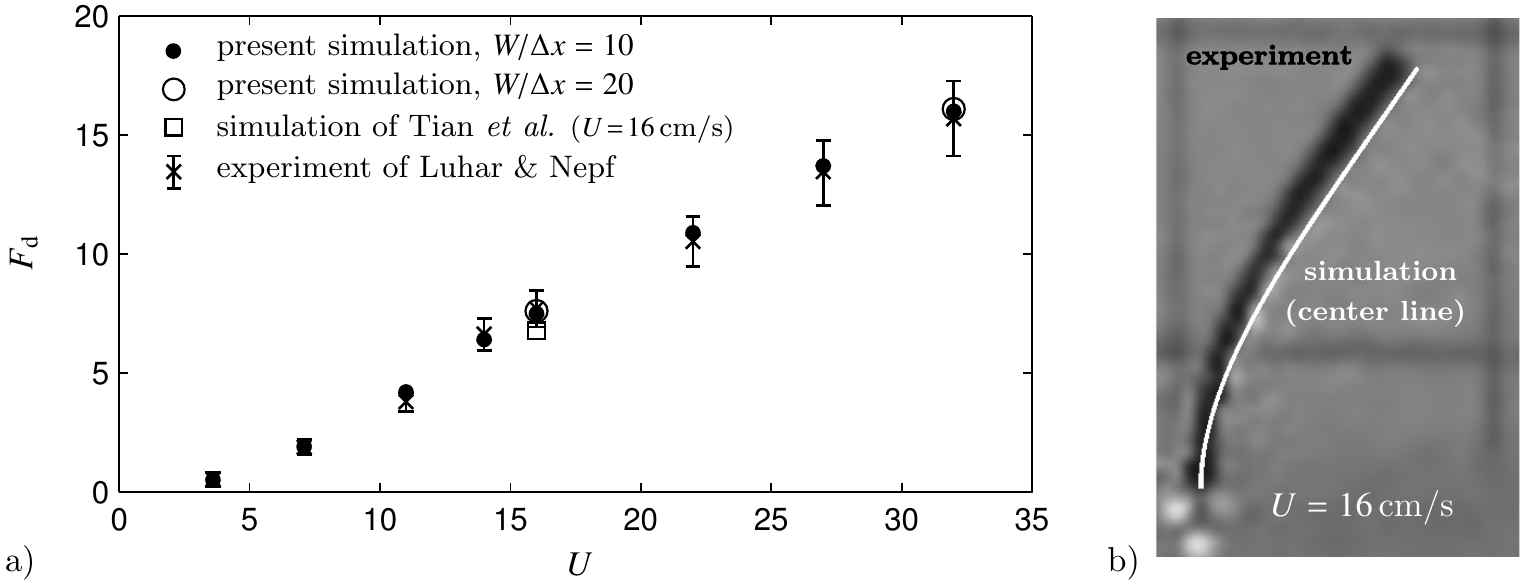}
  \caption{\label{fig:Luhar_3}Results for the rod in cross flow. a) Drag force $F_\mathrm{d}$ against bulk velocity $U$. The present results are compared with the experimental data of Luhar \& Nepf~\cite{Luhar2011} and the simulation results of \text{Tian~\textit{et~al.}~\cite{Tian2014}}. b) Comparison of the deflected rod shape between the experiment and the present simulation at a bulk velocity of $U = \SI{16}{\centi m/s}$. The dark blurred line is the average shape in the experiment, the white line shows the simulation result.}
\end{figure}

The configuration shown in Figure~\ref{fig:Luhar} was experimentally studied by Luhar and Nepf~\cite{Luhar2011} and is well suited as a benchmark problem for steady-state fluid-structure interactions. In contrast to the previous two benchmarks the interaction is pressure dominated as the blade is oriented perpendicular to the mean flow. To provide a uniform flow over the entire length of the rod, it is positioned above the boundary layer of the channel bottom. In the experiment this is realized with the aid of a thin steel rod. The latter is connected to a load sensor which simultaneously measures the integral hydrodynamic force acting on the rod. The structure responds by a large almost steady deflection to the applied fluid load and the wake generated past the rod is three-dimensional and turbulent (Fig.~\ref{fig:Luhar_1_contour}).\\
The laboratory flume used in the experiment is approximated here by a bounded rectangular fluid domain of size $[0;L_x]\times[0;L_y]\times[0;L_z]$ with $L_x=\SI{20}{\centi m}$, $L_y=\SI{16}{\centi m}$ and $L_z=\SI{16}{\centi m}$. Within the domain, the lower end of the rod is positioned at \text{$\vec{c}(Z\!=\!0) = (\SI{5}{\centi m},\SI{5}{\centi m},\SI{8}{\centi m})^\mathrm{T}$}. Tests with different domain sizes have shown that this domain is sufficiently large and does not affect the flow around the rod by boundary effects. At the four lateral boundaries of the domain a free-slip rigid lid condition is applied. The inlet velocity is set to a constant bulk velocity $U$, while a convective outflow condition is used at the outlet. The fluid domain is discretized by $N_x \times N_y \times N_z = 200 \times 160 \times 160$ cells in total, which corresponds to $W/\Delta{x}=10$ grid cells over the width of the structure. In addition, a finer resolution of $W/\Delta{x}=20$ is used to verify grid independence. The rod is discretized by $N_\mathrm{e} = 20$ elements, while $N_\mathrm{e} = 40$ is employed for the fine resolution. A constant time step size of was chosen, yielding $\CFL \approx 0{.}5$. This, for example, results in a time step size of $\Delta{t}=\SI{1e-3}{s}$ for a bulk velocity of $U=\SI{16}{\centi m/s}$ and a grid resolution of $W/\Delta{x}=10$.\\ 
To validate the FSI-solver over a wide range of Reynolds numbers $Re_L=UL/\nu_\f$, simulations were carried out for 8 bulk velocities ranging from $U = \SI{3{.}6}{\centi m/s}$ ($Re_L = 360$) up to $U = \SI{32}{\centi m/s}$ ($Re_L = 3200$). Figure~\ref{fig:Luhar_3} shows a comparison between the present simulation results and the experimental data of Luhar and Nepf~\cite{Luhar2011} over the entire range of bulk velocities~$U$. In addition, the results are compared with a similar IBM simulation carried out by Tian~\textit{et al.}~\cite{Tian2014} for $U = \SI{16}{\centi m/s}$.
\begin{table}[!tb]
  \centering
  \setlength{\tabcolsep}{3pt}
  \begin{tabular}{p{0.027\txtwt} p{0.09\txtwt}|P{0.09\txtwt}|P{0.09\txtwt}|P{0.09\txtwt}|P{0.09\txtwt}|P{0.12\txtwt}|P{0.09\txtwt}|P{0.09\txtwt}|P{0.15\txtwt}}
	  {$U$} &(\SI{}{cm/s})	& 3{.}6 & 7{.}1 & 11 & 14 & 16 & 22 & 27 & 32 \\ \hline
	  {$F_\mathrm{d}$} &(\SI{}{mN}) & 0{.}5 & 1{.}9 & 4{.}3 & 6{.}4 & $7{.}5 \;\, (7{.}6)$ & 10{.}9 & 13{.}7 & $16{.}0 \;\, (16{.}1)$ \\
  \end{tabular}
 \caption{\label{tab:Luhar_3}Average drag force $F_\mathrm{d}$ at different bulk velocities $U$ ranging from \text{$U = \SI{3{.}6}{cm/s}$} up to $U = \SI{32}{cm/s}$. For the simulations performed over the entire range of~$U$\! a grid resolution of $W/\Delta{x}=10$ was used. To verify convergence of $F_\mathrm{d}$, a finer grid resolution of $W/\Delta{x}=20$ was employed at $U = \SI{16}{\centi m/s}$ and $U = \SI{32}{\centi m/s}$ as well (values in brackets).}
\end{table}
With regard to the average drag force $F_\mathrm{d}$, the present results show an excellent agreement with the experimental data over the entire range of bulk velocities $U$. Small differences can be observed for the deflection shape of the rod, shown in Fig.~\ref{fig:Luhar_3}b for a selected velocity of \text{$U = \SI{16}{\centi m/s}$}. Compared to the experimental observation, the rod is slightly more deflected in the simulation. One possible reason can be assigned to the values of the material parameters provided in~\cite{Luhar2011}. Measurement uncertainties of more than 10\% may be expected for the Young modulus $E_\s$ and the density~$\rho_\s$ of the foam material. An additional source of the deviations obtained can be related to the isotropic linear-elastic constitutive relations applied here to simulate a rod made out of non-isotropic foam material. Despite these minor uncertainties in the properties of the experimental setup, the present results show reasonably good agreement with the reference, thus providing another validation of the approach.

\subsubsection{Flow through artificial canopy}
\label{sec:canopy}

\begin{figure}[!ht]
  \setlength{\unitlength}{1cm}
	
  {\begin{minipage}{0.45\txtw}

    \begin{tabular}{p{0.50\txtw} l}
     \hline
     $g = \SI{9{.}81}{m/s^2}$                         & grav. acceleration \\[0.3cm]
     \multicolumn{2}{l}{fluid properties (open water channel):} \\
     $H = \SI{17}{\centi m}$                          & channel height \\
     $\rho_\f = \SI{1000}{kg/m^3}$                    & fluid density \\
     $\nu_\f = \SI{1e-6}{m^2/s}$                      & kin. viscosity \\
     $U = \SI{0.2}{\m\per\s}$                         & bulk velocity \\[0.3cm]
     \multicolumn{2}{l}{structure properties (OHP slides):} \\
     $L = \SI{70}{\milli m}$                          & rod length \\
     $W = \SI{8}{\milli m}$                           & rod width \\
     $T = \SI{0.1}{\milli m}$                         & rod thickness \\
     $\Delta S = \SI{32}{\milli m}$                   & rod spacing \\
     $\rho_\mathrm{s} = \SI{1400}{kg/m^3}$            & structure density \\
     $E_\s = \SI{4{.}8e9}{N/m^2}$                     & Young modulus \\[0.3cm]
     \multicolumn{2}{l}{dimensionless quantities:} \\
     $Re_H = 42000$                                   &	Reynolds number \\
     $\rho_\mathrm{s}/\rho_\f = 1{.}4$                & density ratio \\
     $Ca \approx 17$                                  & Cauchy number \\
     \hline
    \end{tabular}

  \end{minipage}}
  {\begin{minipage}{0.55\txtw}
    \centering
    \includegraphics[trim=0 0 0 0, clip, scale=1]{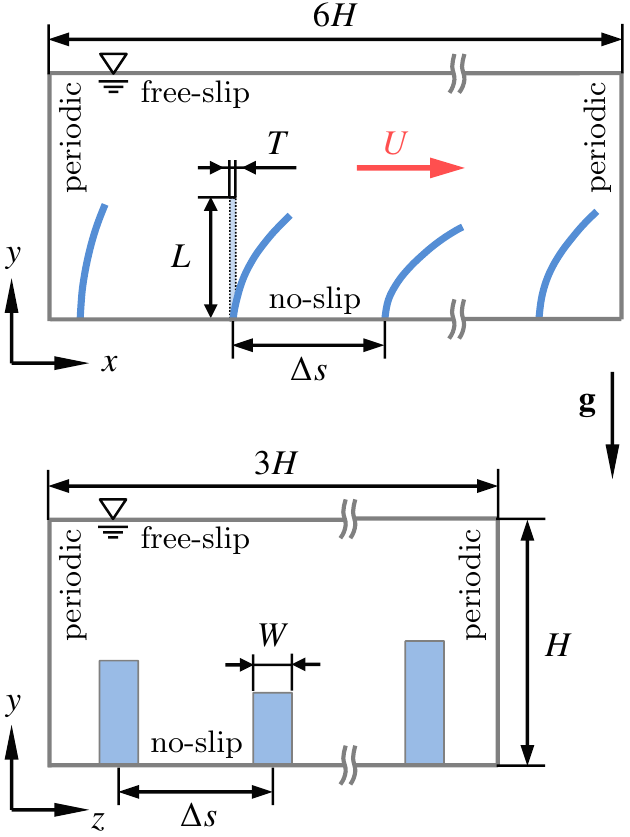}
  \end{minipage}}
  
	\caption{\label{fig:Okamoto_config}Numerical setup of a submerged canopy modeled as an array of flexible rectangular rods made out of polyester overhead projector (OHP) slides with fixation on the bottom plate (same spacing $\Delta S$ in $x$ and $z$-direction), corresponding to the experimental setup of \cite{Okamoto2010}.}
\end{figure}
\begin{figure}[!tb]
	\centering
	\includegraphics[trim=0 0 0 0, clip, scale=1]{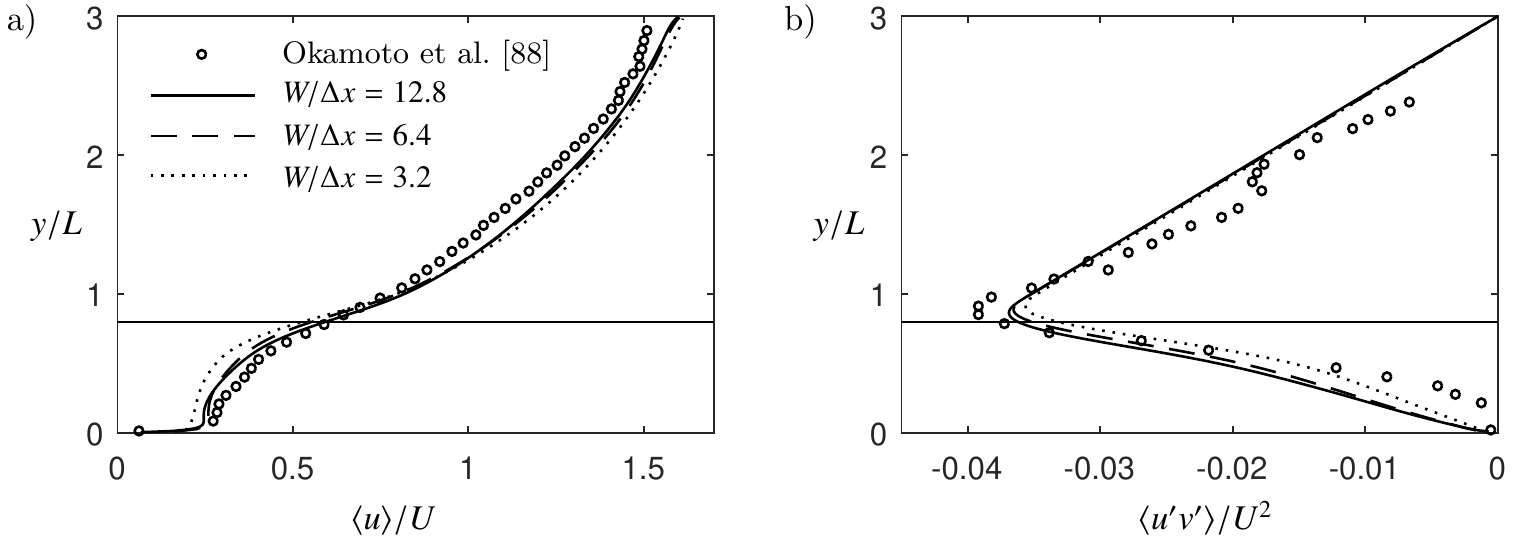}
	\caption{\label{fig:Okamoto_results}Statistical results for the canopy test case in comparison to the experiment~\cite{Okamoto2010} for different grid resolutions. a) Normalized averaged velocity profile $\langle u \rangle/ U $ and b) Reynolds stresses $\langle u'v' \rangle/U^2$. The temporally and spatially averaged height of the canopy with the blades being deflected by the flow $\langle h_c \rangle = 0.8 L$, and represented by a straight horizontal line.}
\end{figure}
\begin{figure}[!tb]
	\centering
	\includegraphics[trim=0 0 0 0, clip, scale=1]{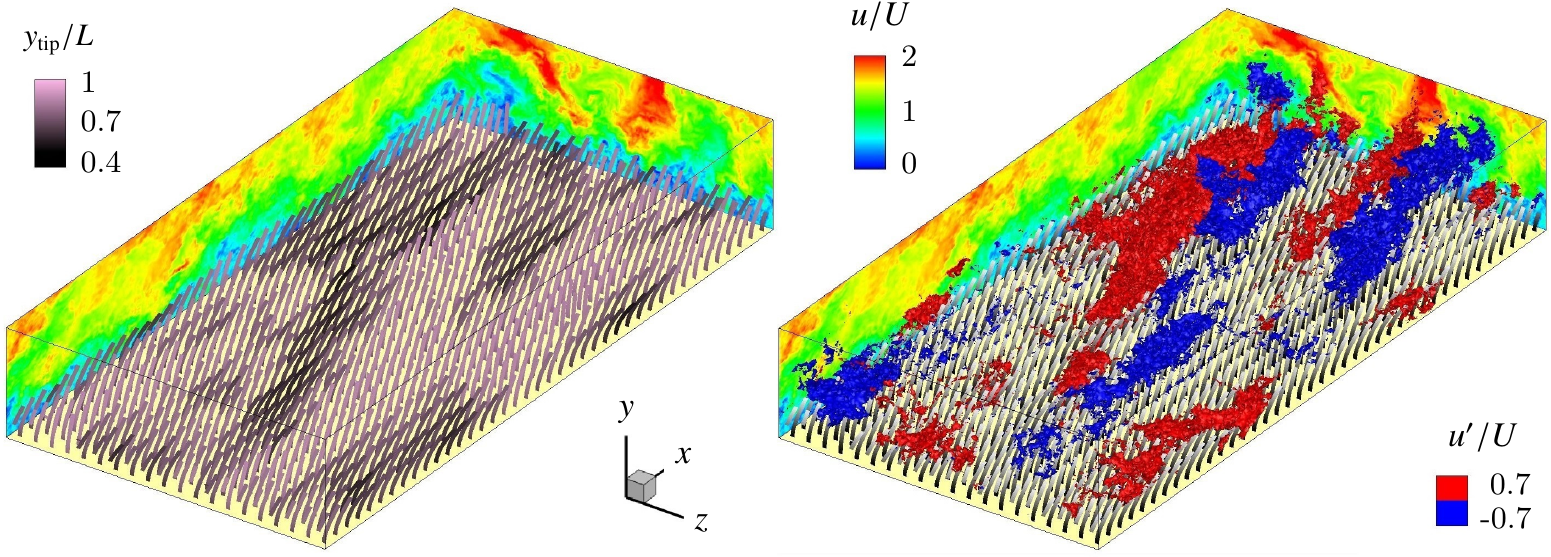}
	\caption{\label{fig:Okamoto_ufluc}LES of a shallow submerged aquatic canopy in a turbulent channel flow, corresponding to the experimental setup of \cite{Okamoto2010}. Both visualizations show the instantaneous, streamwise velocity component $u/U$ in the vertical planes $z=0$ and $x=6H$, at the same arbitrary instant in time. In addition, the left figure shows the array of deflected blades colored with the respective normalized tip elevation ${y_{\mathrm{tip}}/L}$. While some groups of blades are deflected by up to 50\% of the blade length, other groups stand up quite vertically. In the right figure regions of positive and negative velocity fluctuation $u'=u-\langle u \rangle = \pm 0{.}7 U$ are highlighted by iso-surfaces in red and blue, respectively.}
\end{figure}

The final setup addressed here demonstrates the ability of the proposed method for simulations with a large number of highly flexible slender structures in turbulent flow. To validate the FSI-solver for this type of configuration the setup of the experimental work of \cite{Okamoto2010} was simulated as described in Fig.~\ref{fig:Okamoto_config}. The dimension of the computational domain is $6H \times H \times 3H$ in \text{$x$-,~$y$-,~$z$-direction}. It is discretized by cubic cells of size $\Delta x = \SI{0.625}{\milli\m}$, i.e. $W/\Delta x~=~12.8$ grid cells over the blade width. This yields $700$ million grid cells which is at the very edge of what is technically feasible, since the instantaneous flow has to be simulated over a certain duration to be developed and to accumulate statistics. To model the subgrid scale a Smagorinsky constant of $C_s=0.15$ was chosen, as already employed by~\cite{Okamoto2010b} for an LES of canopy flows over rigid blades. The $800$ equally distributed strip-shaped flexible blades are discretized by $30$ elements each in longitudinal direction. The time step was automatically adjusted to yield a $\CFL{}$ number of $0{.}5$. The flow is driven by a spatially constant volume force which is dynamically adjusted in time to maintain a constant bulk velocity of $U=\SI{0{.}2}{m/s}$.  While a no-slip condition is applied at the bottom wall the water surface is approximated by a free-slip rigid lid condition. All remaining boundaries are periodic.\\
For the present set of parameters, in a few cases two or more rods collide. This is taken into account by an own constraint-based collision model, tailored to the properties of Cosserat rods~\cite{Tschisgale2017_coll}.\\
The simulation results for the mean velocity profile $\langle u \rangle/ U $ and the Reynolds stress $\langle u'v' \rangle/U^2$ are given in Fig.~\ref{fig:Okamoto_results} and are compared to the experimental data provided in~\cite{Okamoto2010}. To examine their sensitivity with respect to the grid resolution employed, simulations with coarser resolutions were performed, also included in Fig.~\ref{fig:Okamoto_results}.\\
The comparison to the experimental data of~\cite{Okamoto2010} shows that the mean velocity component $\langle u\rangle$ is slightly underestimated inside the canopy region, while, for reasons of continuity, it is slightly overestimated above the canopy in the free flow region. In this region the Reynolds shear stress $\langle u'v' \rangle/U^2$ has to vary linearly with $y/L$ due to the mean momentum balance. This is very well met by the simulation data. The experimental values, however, exhibit considerable scatter, which might be due to measurement uncertainties or a small amount of averaging. Bearing in mind this issue, together with the known difficulty of precisely determining material properties of the blades, the comparison between experiment and simulation is quite satisfactory.
As demonstrated in \cite{Okamoto2010b} the monami phenomenon (mo=aquatic plant, nami=wave) can be observed for the present set of parameters. It is characterized by a strong interaction between coherent vortices and organized wavelike plant deflection \citep{Okubo2001}. The present simulation shows these well-separated regions of different blade deflection very nicely as they travel through the canopy (Fig.~\ref{fig:Okamoto_ufluc}, a). A deeper analysis of the data reveals that these regions are accompanied by separated longer streaks in streamwise direction of positive and negative velocity fluctuations $u'=u-\langle u \rangle$ (Fig.~\ref{fig:Okamoto_ufluc}, b).
For negative fluctuations $u'<0$ the resulting decreased drag yields more erect blades, while for $u'>0$, in turn, the blades are more deflected. These regions are usually termed low-speed velocity streaks and high-speed velocity streaks, respectively. In general, with experimental methods the shape and temporal evolution of such three-dimensional coherent structures of the flow field are difficult to measure, especially in the canopy region due to the optical inaccessibility resulting from the dense arrangement of moving plants. Especially for such configurations the proposed numerical method is a well-suited means to assess the interaction of numerous highly flexible slender structures with a turbulent flow. Its efficiency can be illustrated by some information on the computing time. The simulation described in this section was undertaken for $44.5$ physical bulk time units $H/U$. This required $\SI{619000}{h}$ CPU time in total. The run was performed on $1024$ Intel processors of different architecture (Intel Westmere, Sandy Bridge, Haswell). The flow solver employing PETSc~\cite{petsc1} and Hypre~\cite{Hypre1} took 93.7\%. Computing the motion of the structures 5.7\% and the coupling by means of the IBM 0.6\%. The present paper is devoted to description and assessment of the numerical method. An in-depth physical analysis is beyond this scope and will be presented elsewhere.

\section{Conclusions}
\label{cha:conclusions}

In the present work, a numerical method suited for fluid-structure interactions of large numbers of slender flexible rods in turbulent flow was developed. The underlying physical model was tailored to this kind of FSI using appropriate model assumptions and simplifications. While the fluid flow is modeled as usual by the three-dimensional Navier-Stokes equations, the motion of the slender structures is described by a powerful one-dimensional rod model, the geometrically exact Cosserat rod model. The basic fluid solver was adopted from the in-house code PRIME~\cite{Kempe2011_diss,Kempe2012}. Concerning the structure solver, the associated Cosserat rod equations were implemented according to recommendations of \text{Lang~\textit{et al.}~\cite{Lang2011}}, who proposed a performance-optimized variant. To describe the interaction of fluid and structure, a new semi-implicit coupling scheme was developed based on an IBM with continuous forcing. It combines the stability of monolithic methods with advantages of partitioned approaches, such as computational efficiency. As a special feature, the coupling is exempt from any global iteration between the fluid part and the structure part, usually performed to ensure numerical stability of partitioned FSI-solvers. In contrast to other non-iterative coupling strategies, the main idea is not based on a stabilization technique, such as relaxation, but on a semi-implicit integration of the coupling terms in the structural equations of motion. It is referred to as semi-implicit coupling here, since only those coupling quantities are treated implicitly which have an effect on the stability of the time integration. Excluded from this are structure positions, constituting the fluid-structure interface. As a result, the proposed coupling scheme requires only a single bidirectional exchange of information between the fluid solver and the structure solver, so that the computing time per individual Cosserat rod could be reduced to a minimum. This makes the developed numerical method highly efficient and particularly suitable for large-scale configurations with a very large number of deformable rods. The method was successfully validated for various test cases with single elastic rods in flow, including the benchmark of Ramm and Wall~\cite{Wall1998, Wall1999} and a three-dimensional setup of a flexible blade in cross flow according to an experiment of Luhar and Nepf~\cite{Luhar2011}. Finally, the proposed method was applied to the flow through an artificial aquatic canopy, consisting of 800 flexible rods, according to the experimental setup of Okamoto and Nezu~\cite{Okamoto2010}. This demonstrates the ability of the present numerical approach for configurations with a large number of slender structures in turbulent flow, and how the generated three-dimensional flow data can be used to gain fundamental insights into the physical of such kind of fluid-structure interactions.

\subsection*{Acknowledgements}
This project was partially founded by DFG under FR 1593/14-1. Computation time was provided by ZIH, Dresden.

\appendix

\section{Proposed FSI algorithm in condensed form}
\label{sec:FSI algorithm}

This section provides an overview of the proposed semi-implicit direct forcing IBM for the coupling of an arbitrary number of Cosserat rods to the Navier-Stokes equations. All relevant equations are given for an individual Runge-Kutta sub-step $r$ within the time interval \text{$t\in[t^{r-1},t^r]$}. With the three-step Runge-Kutta scheme employed here, each time step $\Delta t$ consists of three sub-steps, so that quantities at the new time level $t^{n+1}$ are provided after the third sub-step. Quantities of the previous time level $t^n$ are denoted by superscript 0, e.g. the velocity field $\vec{u}^0(\vec{x}_{ijk})$. The $r$th Runge-Kutta sub-step can be summarized as follows:

\begin{subequations}
\setlength{\abovedisplayskip}{0pt}
\begin{equation}
	\label{alg:fsi_solve_01}
  \frac{ \tilde{\vec{u}} - \vec{u}^{r-1}}{\Delta t} =  2\alpha_{r}\: \nu_\f \:\nabla^{2}\vec{u}^{\:r-1}
  - 2\alpha_{r} \: \nabla \!\left(p^{\:r-1}\!/\rho_\f\right) - \gamma_{r} \:\nabla \cdot \left(\vec{u}\otimes\vec{u} \right) ^{\:r-1}
  - \zeta_{r} \: \nabla \cdot \left(\vec{u}\otimes\vec{u} \right) ^{\:r-2} + \vec{f}_V
\end{equation}

\begin{equation}
	\label{alg:fsi_solve_02}
	 \vec{\xi}_l^{r-1} \!= {\vec{q}}^{r-1}_e\!\qm\vec{\xi}_{\text{\tiny{0}},l}\qm\bar{\vec{q}}^{r-1}_e \qquad \text{and} \qquad \vec{x}_l^{r-1} \!= \vec{c}^{r-1}_{e} \!\!+ \vec{\xi}_l^{r-1}
\end{equation}

\begin{equation}
	\label{alg:fsi_solve_03}
	\tilde{\vec{u}}(\vec{x}_l^{r-1})  = \sum^{N_x}_{i=1} \sum^{N_y}_{j=1} \sum^{N_z}_{k=1}\tilde{\vec{u}}(\vec{x}_{ijk}) \:\delta_h(\vec{x}_{ijk}-\vec{x}_l^{r-1}) \:h^3
\end{equation}

\begin{equation}
	\label{alg:fsi_solve_04}
	\ddot{\vec{c}}_{e+\frac{1}{2}} = \,\frac{1}{\rho_\s A}\,\left\{\,\left(\vec{q}\qm\fint\tra\!\qm\bar{\vec{q}}\right)' + \fext_\IBM\,\right\}_{e+\frac{1}{2}} \!+\, \Big(1-\rho_\mathrm{f}/\rho_\mathrm{s}\Big)\,\vec{g}
\end{equation}

\begin{equation}
	\label{alg:fsi_solve_05}
	\ddot{\vec{q}}_e = \frac{1}{2\rho_\s}\,\mat{\mathcal{M}}_e\cdot\left\{\,4\rho_\s \,\dot{\vec{q}}\qm\mat{\mathcal{I}\tra}\!\cdot\!\left(\dot{\bar{\vec{q}}}\qm\vec{q}\right) + \vec{c}'\!\qm\vec{q}\qm\fint\tra + \left(\vec{q}\qm\mint\tra\right)' + \vec{q}'\!\qm\mint\tra + \mext_\IBM\qm\vec{q} \,\right\}_e - \|\dot{\vec{q}}_e\|^2\vec{q}_e
\end{equation}

\begin{equation}
	\label{alg:fsi_solve_14}
	\vec{q}^r_e \leftarrow \vec{q}^r_e/\|\vec{q}^r_e\| \qquad \text{and} \qquad \dot{\vec{q}}^r_e \leftarrow \dot{\vec{q}}^r_e- (\vec{q}^r_e\!\cdot\!\dot{\vec{q}}^r_e)\,\vec{q}^r_e
\end{equation}

\begin{equation}
	\label{eqn:fsi_solve_15}
	\vec{u}^{r}_\Gamma(\vec{x}_l^{r-1}) = \dot{\vec{c}}^r_{e} + \left(2\,\dot{\vec{q}}^r_e\!\qm\bar{\vec{q}}^r_e\right) \times \vec{\xi}_l^{r-1}
\end{equation}

\begin{equation}
	\label{alg:fsi_solve_16}
	\bar{\vec{f}}_\IBM(\vec{x}_l^{r-1}) = \frac{\vec{u}^{r}_\Gamma(\vec{x}_l^{r-1}) - \tilde{\vec{u}}(\vec{x}_l^{r-1})}{2\alpha_r\,\Delta t}
\end{equation}

\begin{equation}
	\label{alg:fsi_solve_17}
	\bar{\vec{f}}_\IBM(\vec{x}_{ijk}) = \sum_{\;\vec{x}_{l}\,\in\,\Gamma_e} \bar{\vec{f}}_\IBM(\vec{x}_l^{r-1})\: \delta_h(\vec{x}_{ijk}-\vec{x}\vec{}_l^{r-1}) \:\Delta V_l
\end{equation}

\begin{equation}
	\label{alg:fsi_solve_18}
	\nabla^{2}\vec{u}^{*} - \frac{\vec{u}^{*}}{\alpha_{r} \nu_\f \:\Delta t} = \nabla^{2}\vec{u}^{r-1} 
	- \frac{ \tilde{\vec{u}} + 2\alpha_r\,\Delta t\:\bar{\vec{f}}_\IBM }{\alpha_{r} \nu_\f \:\Delta t}
\end{equation}

\begin{equation}
	\label{alg:fsi_solve_19}
	\nabla^{2} \phi^{r} = \nabla \cdot \vec{u}^{*}
\end{equation}

\begin{equation}
	\label{alg:fsi_solve_20}
	\vec{u}^{r} = \vec{u}^{*} - \nabla \phi^{r}
\end{equation}

\begin{equation}
	\label{alg:fsi_solve_21}
	\frac{p}{\rho_\f}^{\! r} = \frac{p}{\rho_\f}^{\! r-1} + \frac{\phi^{r}}{ 2\:\alpha_{r}\:\Delta t} - \frac{\nu_\f}{2} \:\nabla^{2} \phi^{r}
\end{equation} 
\end{subequations}\\

The remaining terms contained in the Cosserat rod equations of motion~\eqref{alg:fsi_solve_04}, \eqref{alg:fsi_solve_05} are:
\begin{subequations}
\begin{flalign}
	\label{alg:fsi_solve_06}
	\!\left(\vec{q}\qm\fint\tra\!\qm\bar{\vec{q}}\right)'_{e+\frac{1}{2}} \!\!\!= \big(\vec{q}_{e+1}\qm\fint_{\text{\tiny{0}},e+1}\!\qm\bar{\vec{q}}_{e+1} - \vec{q}_{e}\qm\fint_{\text{\tiny{0}},e}\!\qm\bar{\vec{q}}_{e}\big)\,/\,\Delta Z &&
\end{flalign}
\begin{flalign}
	\label{alg:fsi_solve_08-2}
	\fext_{\IBM,e+\frac{1}{2}} = (\fext_{\IBM,e} + \fext_{\IBM,e+1})/2 &&
\end{flalign}
\begin{flalign}
	\dot{\vec{c}}_{e} = (\dot{\vec{c}}_{e+\frac{1}{2}} + \dot{\vec{c}}_{e-\frac{1}{2}})/2 &&
\end{flalign}
\begin{flalign}
	\label{alg:fsi_solve_07}
	\mat{\mathcal{M}}_e = \mat{\mathcal{Q}}(\vec{q}_e)\cdot\mat{\mathcal{I}}^{-1}\tra\cdot\mat{\mathcal{Q}}^\top\!(\vec{q}_e) \qquad \text{with} \qquad \mat{\mathcal{Q}}(\vec{q}) = 
	\begin{pmatrix} \Re(\vec{q}) & -\Im(\vec{q}) \\ \Im(\vec{q}) & \Re(\vec{q})\,\mat{\mathbb{I}}+[\Im(\vec{q})]_\times \end{pmatrix} &&
\end{flalign}
\begin{flalign}
	\vec{c}'_e = \big(\vec{c}_{e+\frac{1}{2}}-\vec{c}_{e-\frac{1}{2}}\big)\,/\,\Delta Z &&
\end{flalign}
\begin{flalign}
	\label{alg:fsi_solve_08}
	\!\left(\vec{q}\qm\mint\tra\right)'_e + \vec{q}'_e\!\qm\mint\trae = \big(\vec{q}_{e+1}\!\qm\mint_{\text{\tiny{0}},e+\frac{1}{2}} - \vec{q}_{e-1}\!\qm\mint_{\text{\tiny{0}},e-\frac{1}{2}}\big)\,/\,\Delta Z &&
\end{flalign}
\begin{flalign}
	\label{alg:fsi_solve_09}
	\fint\trae = \,\left[ \mat{C}_\gamma\cdot\left(\vec{\gamma}\tra\!-\vec{\gamma}\tra|_{t=0}\right) \,+\, 2\,\mat{C}_{\dot{\gamma}}\cdot\dot{\vec{\gamma}}\tra \right]_e
	\quad \text{with} \quad \vec{\gamma}\trae\!= \bar{\vec{q}}_e\qm\vec{c}'_e\qm\vec{q}_e &&
\end{flalign}
\begin{flalign}
	\label{alg:fsi_solve_10}
	\mint_{e+\frac{1}{2}} = \Big[ \mat{C}_\kappa\cdot\left(\vec{\kappa}\tra\!-\vec{\kappa}\tra|_{t=0}\right) \,+\, 2\,\mat{C}_{\dot{\kappa}}\cdot\dot{\vec{\kappa}}\tra \,\Big]_{e+\frac{1}{2}}
	\quad \text{with} \quad \vec{\kappa}_{\text{\tiny{0}},e+\frac{1}{2}}\!= \frac{2\sqrt{2}}{\Delta Z}\,\frac{\Im\big(\bar{\vec{q}}_e\!\qm\vec{q}_{e+1}\big)}{\sqrt{1+\Re\big(\bar{\vec{q}}_e\!\qm\vec{q}_{e+1}\big)}} \quad \text{(see \cite{Lang2011})} &&
\end{flalign}
\begin{flalign}
	\label{alg:fsi_solve_11}
	\fext_{\IBM,e}\,\Delta Z = -\left[ \frac{{\vec{p}_\Gamma} - \vec{p}^{r-1}_\Gamma}{t-t^{r-1}} + \frac{\vec{p}^{r-1}_\Gamma - \tilde{\vec{p}}}{2\alpha_{r} \:\Delta t} \right]_e \quad \text{with} 
	\quad &\vec{p}_{\Gamma,e} = \left[ m\,\dot{\vec{c}} + 2\dot{\vec{q}}\qm\bar{\vec{q}}\times\vec{q}\qm\vec{s}\tra\!\qm\bar{\vec{q}} \,\right]_e &&\\
	\notag
	&\tilde{\vec{p}}_e = \sum_{\;\vec{x}_{l}\,\in\,\Gamma_e} \Delta m_l\,\tilde{\vec{u}}(\vec{x}_{l}^{r-1}) &&
\end{flalign}
\begin{flalign}
	\label{alg:fsi_solve_12}
	\mext_{\IBM,e}\,\Delta Z = -\left[ \frac{{\vec{l}_\Gamma} - \vec{l}^{r-1}_\Gamma}{t-t^{r-1}} + \frac{\vec{l}^{r-1}_\Gamma - \tilde{\vec{l}}}{2\alpha_{r} \:\Delta t} \right]_e \quad \text{with} \quad &\vec{l}_{\Gamma,e} = \left[ \vec{q}\qm\vec{s}\tra\!\qm\bar{\vec{q}}\times\dot{\vec{c}} + \vec{q}\qm\mat{\mathcal{J}}_\text{\tiny{0}}(2\bar{\vec{q}}\qm\dot{\vec{q}})\qm\bar{\vec{q}} \, \right]_e &&\\
	\notag
	&\tilde{\vec{l}}_e = \sum_{\;\vec{x}_{l}\,\in\,\Gamma_e} \Delta m_l\,\vec{\xi}^{r-1}_{l}\times\tilde{\vec{u}}(\vec{x}^{r-1}_{l}) &&
\end{flalign}
\begin{flalign}
	\label{alg:fsi_solve_13}
m_e = \sum_{\;\vec{x}_{l}\,\in\,\Gamma_e} \Delta m_l \, , \qquad 
\vec{s}_{\text{\tiny{0}},e} = \sum_{\;\vec{x}_{l}\,\in\,\Gamma_e} \Delta m_l\,\vec{\xi}_{\text{\tiny{0}},l} \, , \qquad 
	\mat{\mathcal{J}}_{\text{\tiny{0}},e} = 0\oplus\!\!\! \sum_{\;\vec{x}_{l}\,\in\,\Gamma_e} \Delta m_l\,[\vec{\xi}_{\text{\tiny{0}},l}]^\top_\times \cdot [\vec{\xi}_{\text{\tiny{0}},l}]_\times \quad. &&
\end{flalign}
\end{subequations}

\newpage
\bibliographystyle{elsarticle-num}
\bibliography{bibliography}

\end{document}